\documentclass[reprint, amsmath, amssymb, aps, superscriptaddress]{revtex4-2}

\usepackage{graphicx}
\usepackage{dcolumn}
\usepackage{bm}
\usepackage[colorlinks,linkcolor=blue,citecolor=blue]{hyperref}

\begin{document}

\preprint{APS/123-QED}

\title{Centrality and Universality in Scale-Free Networks}
\author{V. Adami}
\affiliation{Department of Physics, University of Mohaghegh Ardabili, P.O. Box 179, Ardabil, Iran}
\author{S. Emdadi-Mahdimahalleh}
\affiliation{Department of Electrical and Computer Engineering, University of Akron, Akron, Ohio 44325, USA}
\author{H. J. Herrmann}
\affiliation{Departamento de Física, Universidade Federal do Cear\' a, 60451-970, Fortaleza, Ceará,
	Brazil}
\affiliation{PMMH, ESPCI, 7 quai St. Bernard, 75005 Paris, France}
\author{M. N. Najafi}
\affiliation{Department of Physics, University of Mohaghegh Ardabili, P.O. Box 179, Ardabil, Iran}
\email{morteza.nattagh@gmail.com}
    
\date{\today}

\begin{abstract}
We propose a novel paradigm for modeling real-world scale-free networks, where the integration of new nodes is driven by the combined attractiveness of degree and betweenness centralities, the competition of which (expressed by a parameter $0\le p\le 1$) shapes the structure of the evolving network. We reveal the ability to seamlessly explore a vast landscape of scale-free networks, unlocking an entirely new class of complex networks that we call \textit{stars-with-filament} structure. Remarkably, the average degree $\bar k$ of these networks grows like $\log t$ to some power, where $t$ is time and the average shortest path length grows logarithmically with the system size for intermediate $p$ values, offering fresh insights into the structural dynamics of scale-free systems. Our approach is backed by a robust mean-field theory, which nicely captures the dynamics of 
$\bar{k}$. We further unveil a rich,
$p$-dependent phase diagram, encompassing 47 real-world scale-free networks, shedding light on previously hidden patterns. This work opens exciting new avenues for understanding the universal properties of complex networks.
\end{abstract}

\maketitle

Scale-free networks exhibit power-law scaling in statistical measures~\cite{wikimedia_dumps, kunegis2013konect, auer2007dbpedia, omodei2015characterizing, richters2011trust, cho2011friendship, caida_as_relationships, leskovec2005graphs, fire2013link, maniu2011casting, gursoy2018influence, paranjape2017motifs, de2015muxviz, hu2018molecular, zhang2005collecting, rocha2011simulated, yang2015nationtelescope, dallas2018gauging, fellbaum2010wordnet, alberich2002marvel, fire2014computationally, leskovec2007graph, chess, milo2004superfamilies, klimt2004enron, wachs2021corruption, de2009social, de2015identifying}, the exponents of which can be employed to identify universality classes. Understanding the formation mechanisms of real-world scale-free networks requires recognizing that newly added nodes tend to connect not only to highly interactive hubs with high degree centrality, as in the Barabási-Albert model~\cite{barabasi1999emergence}, but also to nodes distinguished by other types of centralities, such as betweenness centrality, which facilitates access to other parts of the network. An example is the person-country affiliation network~\cite{auer2007dbpedia}, which has a bipartite structure and whose exponents deviate significantly from those predicted by the Barabási-Albert model. This deviation, as we demonstrate in this paper, arises because newcomers prioritize connections based on both betweenness and degree centralities, aiming to balance linking to well-connected super-hubs with broadening access to other regions of the network. We reveal that this interplay between degree and betweenness centralities defines the network's structure, and models must account for both measures to accurately capture the observed phenomena in a wide range of scale-free networks.\\

The number of links connected to a node $k$, and betweenness centrality, defined as~\cite{freeman1977set}
\begin{equation}
b_j=\frac{\sum_{j_1j_2}\sigma_{j_1jj_2}}{\sum_{j_1j_2}\sigma_{j_1j_2}},
\end{equation}
follow distribution functions given by
\begin{equation}
	P(k) \sim k^{-\gamma},\ P(b)\sim b^{-\delta},
\end{equation}
in scale-free networks, where $\gamma$, $\delta$, and $\eta$ are scaling exponents where $b \propto k^{\eta}$. In these relations, $\sigma_{j_1jj_2}$ represents the number of shortest paths between nodes $j_1$ and $j_2$ that pass through node $j$, and $\sigma_{j_1j_2}$ is the total number of shortest paths between $j_1$ and $j_2$. Betweenness centrality plays a critical role in maintaining network stability and facilitating information flow. Scaling relations between $k$ and $b$ yield $\eta = \frac{\gamma - 1}{\delta - 1}$ under certain conditions~\cite{masoomy2023relation}. Network geometry studies reveal connections between these exponents and the geometrical properties of scale-free  networks~\cite{song2005self,PhysRevLett.100.248701}. These scaling laws also offer insights into the average shortest path length, $l$, which in the Barabási-Albert model scales as $\frac{\ln N}{\ln \ln N}$~\cite{fronczak2004average}, where $N$ is the number of nodes. The exponents $\gamma$ and $\eta$ also play key roles in epidemic spreading on networks~\cite{pastor2001epidemic, boguna2002epidemic}.

Goh et al.~\cite{goh2001universal, goh2002classification} conjectured a universal classification for scale-free complex networks based on the betweenness centrality exponent $\delta$, dividing real networks into two universality classes: $\delta = 2.2$ (e.g., protein interaction networks) and $\delta = 2.0$ (e.g., internet and WWW networks). Barthélemy~\cite{barthelemy2003comment} argued instead that $\delta$ varies continuously with $\gamma$ and depends on specific network features. He proposed~\cite{barthelemy2004betweenness} the relationship $\delta \geq \frac{\gamma + 1}{2}$, particularly for scale-free trees, where a mean-field argument obtains for $\eta$ its maximal value of 2. While this applies to some networks, others, like visibility graph networks, violate the conjecture~\cite{masoomy2023relation}. The Barabási-Albert model, consistent with $\gamma = 3$, $\delta = 2$, and $\eta = 2$, satisfies Barthélemy’s conjecture but fails to reproduce the exponents observed in real networks. Generalizations of the Barabási-Albert model incorporate additional factors, such as aging and fitness, to better capture the dynamics of scale-free networks~\cite{albert2000topology, bianconi2001competition, dorogovtsev2000scaling, krapivsky2000connectivity}. This letter focuses on essential questions concerning the mechanism behind the universality of complex networks and the dynamics driving their formation into scale-free structures.\\

\begin{figure}[t]
	\centering
	\includegraphics[scale=0.5]{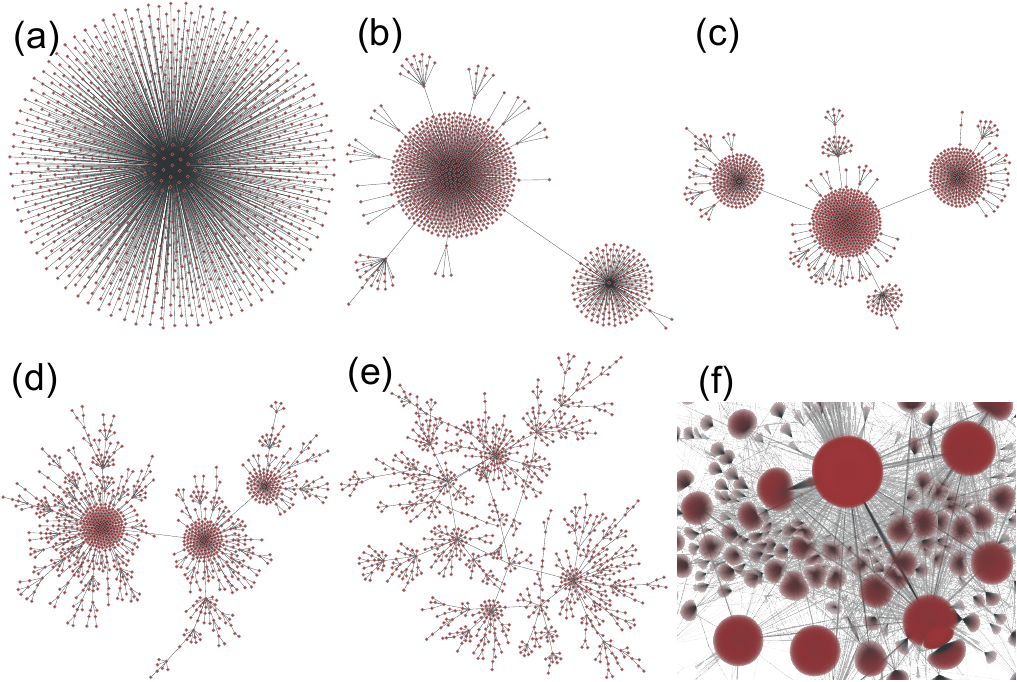}
	\caption{Visualization of the model's networks for $N=1000$ and $p$ values: (a) $0$ (star), (b) $0.03$, (c) $0.1$, (d) $0.5$, and (e) $1$ (BA). (f) shows cropped part of person-country network with 592414 nodes and average degree $\left\langle k \right\rangle \approx 2.15$ which corresponds to effective $p\approx 0.1$~\cite{auer2007dbpedia} taken from \href{https://doi.org/10.5281/zenodo.7839981}{Netzschleuder catalog and repository of network datasets}.}
	\label{fig:model_visualization}
\end{figure}

\begin{figure*}[t] 
  \centering \includegraphics{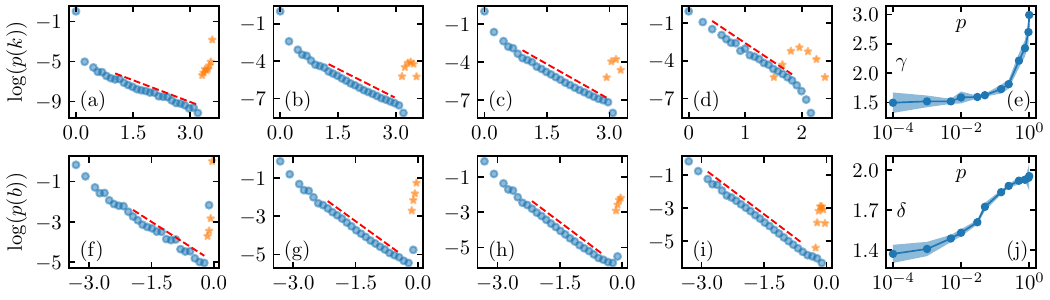}
  \caption{(a), (b), (c), and (d) show the degree distributions, whereas (f), (g), (h), and (i) display the betweenness distributions of the networks generated by our model in the log-log scale, exhibiting power-law behavior. The $p$ values are set to $0.0001$ for (a) and (f), $0.03$ for (b) and (g), $0.15$ for (c) and (h), and $1$ for (d) and (i). The stars represent the degree (betweenness) centrality concentrations of the network's super-hubs, while the filled circles depict the distributions of the remaining nodes. Straight dashed lines indicate the fittings used to determine the degree and betweenness exponents, denoted by $\gamma$ and $\delta$, respectively. (e) and (j) illustrate how the exponents $\gamma$ and $\delta$ vary with changes in the parameter $p$. Here, the network size is $N=4000$ with $m=1$.}
  \label{fig:degree_bet_dist}
\end{figure*}
Our model implements preferential attachment tuned by an external parameter $0 \le p \le 1$, leading us to name it the $p$ centrality-driven attachment ($p$-CDA) model. Let us define a network $\mathcal{G}_p\left\lbrace \mathcal{E}; t \right\rbrace$ with $t$ nodes ($t$ also plays the role of time), where $\mathcal{E}$ is the set of edges between nodes. At each time step, a new node is added to the network, accompanied by a new link connecting it to an existing node. Upon adding the $(t+1)$th node, a new link is formed: we have $\mathcal{G}_p\left\lbrace \mathcal{E}; t \right\rbrace \to \mathcal{G}_p\left\lbrace \mathcal{E}'; t+1 \right\rbrace$, where $\mathcal{E}'$ is the union of $\mathcal{E}$ and a new edge, according to the following rule. The destination node for the new link is selected stochastically, either based on its degree centrality with probability $p$, or on its betweenness centrality with probability $1-p$. More precisely, upon adding a new node, a random number $r \in [0,1]$ is drawn from a uniform distribution. If $r \le p$ (with probability $p$), the destination node is chosen according to the degree centrality of nodes, expressed by the degree centrality based probability measure $\Pi(k_i, t) \equiv \frac{k_i(t)}{\sum_{j=1}^t k_j}$ for the $i$th node; otherwise (with probability $1-p$), the destination node is chosen according to the betweenness centrality of nodes given by the degree centrality based probability measure $\Pi(b_i, t) \equiv \frac{b_i(t)}{\sum_{j=1}^t b_j}$ for the $i$th node. This model acts as an interpolation between two extremes: degree centrality and betweenness centrality based  preferential attachments. Notably, the Barabási-Albert model emerges as the limiting case when $p \to 1$, while for $p = 0$ the network forms a star graph, where all betweenness values are zero except for a dominant hub. We expect that the Barabási-Albert universality class is modified when $p \ne 1$, as $k$ and $b$ are essentially uncorrelated in scale-free networks, meaning a high (or low) $k$ does not necessarily imply a corresponding high (or low) $b$. We sweep the $p$ values to explore the $p$-CDA model. A large $p$ value indicates that nodes are more likely to consider the degree centrality of potential connections, whereas a smaller $p$ value suggests that nodes are more inclined to focus on betweenness centrality when deciding which nodes to connect to.

Our simulations begin with a setup of two nodes linked together. The growth dynamics proceeds according to the algorithm described above. If a new node favors betweenness centrality and the existing nodes have betweenness centrality of zero, we allow that new node to randomly connect to any of the existing nodes. We continue this process until $N$ nodes are present in the system, which defines the system size. We ran the simulations for different $p$ values, with logarithmic increments. Figure~\ref{fig:model_visualization} illustrates how the network evolves as $p$ increases from zero (a star graph) to one (a Barabási-Albert network). For intermediate $p$ values, we observe the appearance of semi-hub nodes, creating a network that is a mixture of super-hubs and other branches. Such a \textit{stars-with-filament} structure has already been observed in~\cite{auer2007dbpedia}, an example of which is shown in Fig.~\ref{fig:model_visualization}f with an effective $p=0.10003$ to be compared with Fig.~\ref{fig:model_visualization}c. The ``stars" refer to the super-hubs and the ``filaments" (growing with increasing $p$) to the rest which depends on the network structure. This evolution is reflected in the distribution functions of degree and betweenness centralities as shown in Fig.~\ref{fig:degree_bet_dist}. We see that for intermediate $k$ and $b$, $p(k)$ and $p(b)$ exhibit power-law behavior for all considered $p \ne 0$ values, indicating the scale-free nature of the network. The data points at the tails (star symbols) are fitted best with a Gaussian distribution, representing the super-hub nodes. From this point onward, we perform separate calculations for the super-hubs and the rest of the nodes, as they follow different statistical behaviors. Figures~\ref{fig:degree_bet_dist}(e) and~\ref{fig:degree_bet_dist}(j) display respectively the $\gamma$ and $\delta$ exponents as functions of $p$ for $N = 4000$. As $p$ decreases, the $\gamma$ exponent smoothly transitions from $3$ (the Barabási-Albert limit) to approximately $1.5$, while $\delta$ decreases from $2$ to about $1.37$. This not only shows that the $p$-CDA model introduces a new paradigm with continuously varying exponents, but also challenges the common belief that $\gamma$ should fall within the interval $[2, 3]$ for scale-free networks.

\begin{figure}[t] 
  \centering
  \includegraphics{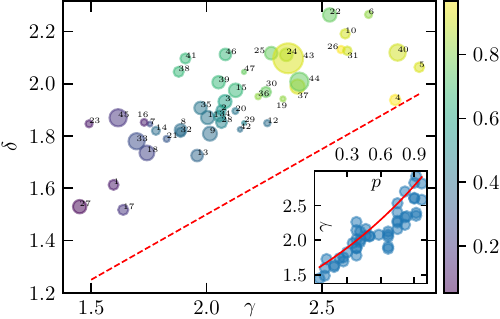}
  \caption{$\delta-\gamma$ phase space of $47$ real networks~\cite{wikimedia_dumps, kunegis2013konect, auer2007dbpedia, omodei2015characterizing, richters2011trust, cho2011friendship, caida_as_relationships, leskovec2005graphs, fire2013link, maniu2011casting, gursoy2018influence, paranjape2017motifs, de2015muxviz, hu2018molecular, zhang2005collecting, rocha2011simulated, yang2015nationtelescope, dallas2018gauging, fellbaum2010wordnet, alberich2002marvel, fire2014computationally, leskovec2007graph, chess, milo2004superfamilies, klimt2004enron, wachs2021corruption, de2009social, de2015identifying} indexed by numbers from 1 to 47 (refer to the Appendix) and shown by disks whose radius is proportional to the errors. The colors show the amount of $p$ (according to the color bar), resulting from the fitting to the $p$-CDA model. The dashed line is Barthelemy's threshold line, $\delta=\frac{\gamma+1}{2}$. The inset shows the resulting $\gamma$ of the real networks in terms of their corresponding $p$ with the red line showing the $p$-CDA model prediction.}
  \label{fig:phase_space}
\end{figure}

To assess the applicability of the $p$-CDA model in reproducing the exponents observed in real-world networks, we analyzed 47 famous scale-free networks. For each of these networks, we extracted the distribution functions of degree and betweenness centralities, and estimated the exponents $\gamma$, $\delta$, and $\eta$ by finding the most appropriate intervals where the networks show scaling relations. See the Appendix and Table~\ref{tab:network_comparison} and Figs.~\ref{fig:sample_RNs} till~\ref{fig:real_networks_9} for detailed information on the types of networks and the data fitting process. Figure~\ref{fig:phase_space} presents the phase space for these real-world networks. The horizontal axis represents $\gamma$, and the vertical axis represents $\delta$, while the radii of the circles are proportional to the errors. To fit the data to the $p$-CDA, we estimated the optimal $p$ by averaging over the three $p$ values extracted from the exponents $\gamma$, $\delta$, and $\eta$ that provided the best fit (see Table~\ref{tab:network_comparison} for details). We observed very good agreement between the model and the real-world data. The colors in Fig.~\ref{fig:phase_space} represent the best-fit $p$ values and we see a continuous gradient of color from left to right. The inset of the figure shows $\gamma$ as a function of $p$, with the red line representing the prediction of the $p$-CDA model, which matches the observed data. Additionally, we examined the loop structure of the real networks and found that most of the considered networks closely resemble trees, similar to the $p$-CDA model. For instance, the clustering coefficient of these networks is either zero or very small. The dashed line represents Barthélemy's conjecture, $\delta_c = \frac{\gamma+1}{2}$, and all the networks adhere to his prediction  $\delta>\delta_c$. The correspondence between the real-world networks and the $p$-CDA model leads us to conclude that betweenness centrality is a crucial factor in the formation of real-world growing networks.

\begin{figure}[t] 
  \centering
  \includegraphics{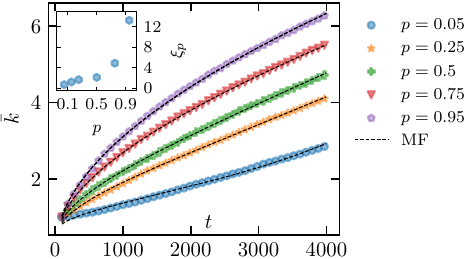}
  \caption{Time evolution of the average degree $\bar{k}$ for a node introduced at time step 100 in the $p$-CDA model, averaged over 1500 realizations. Dashed lines indicate mean-field predictions from Eq.~\ref{Eq:averageK}. The inset shows the dependence of the exponent $\xi_p$ on $p$. The error bars are smaller than the symbol sizes.}
  \label{fig:mean_field}
\end{figure}

To better understand the properties of the $p$-CDA model, we develop a mean-field theory description based on the growth rate of the degree centrality of a typical node. The growth of the $i$th node is given probabilistically as:
\begin{equation}
\frac{dk_i}{dt}= p\Pi(k_i,t)+(1-p)\Pi(b_i,t).
\end{equation}
which states that the growth of $k_i$, the degree of node $i$, is determined by $\Pi(k_i,t)$ with probability $p$, and by $\Pi(b_i,t)$ with probability $1-p$. For more details about this mean-field scheme, refer to the section~\ref{sec:mean_filed_appendix} in the Appendix. For $p$-CDA networks the proportionality parameter of the scaling relation between $b$ and $k$ is generally time-dependent, i.e. $b = A_{bk}^{(p)} t^{-\tau_p}k^{\eta_p}$, where $A_{bk}^{(p)}$ is a proportionality constant which depends on $p$, and $\tau_p$ is an exponent. Our numerical observations suggest that $\tau_p=1$ for all $p$ values. It is essential to 
decompose the summation of betweenness to two parts: $\sum_j b_j(t)=b^{\text{Hub}}_p(t)+\sum'_{j}b_{j}(t)$, where $b^{\text{Hub}}_p$ is the betweenness centrality of the super-hub, and $\sum'$ is the summation over all nodes except the super-hub. For sufficiently large times, $b^{\text{Hub}}_p(t)$ becomes a $p$-dependent and $t$-independent parameter parameter $ \bar{b}^{\text{Hub}}_p$, while $\sum'_{j}b_{j}(t)$ grows smoothly with time in a gentle way. The solution is a complicated one for the general case. However, given that the residual betweenness $\sum'_jb_j(t)$ is a gentle and slow-varying function of time, if we substitute it by an average constant residual betweenness ($ \bar{b}_p^r$) for the intermediate times we find
\begin{equation}
\frac{dk_i}{dt}=c_p^{(1)}\frac{k_i}{t}+c_p^{(2)}\frac{k_i^{\eta_p}}{t}.
\end{equation}
where $c_p^{(1)}\equiv \frac{p}{2}$ and $c_p^{(2)}\equiv \frac{A_{bk}^{(p)}(1-p)}{\bar{b}^{\text{Hub}}_p+\bar{b}_p^r}$. We observe that $\eta_p > 1$ for all $p$ values, which implies that the term $k_i^{\eta_p}$ dominates for sufficiently large $k_i$ values. This leads to:
\begin{equation}
\bar{k}=\left(a_p-b_p\log t\right)^{-\xi_p},
\label{Eq:averageK}
\end{equation}
where $a_p\equiv \frac{1}{k_0^{\eta_p-1}}+(\eta_p-1)c_p^{(2)}\log t_0$, $b_p\equiv (\eta_p-1)c_p^{(2)}$, and $\xi_p\equiv \frac{1}{\eta_p-1}$. Here, $k_0$ and $t_0$ are initial values. This demonstrates that our model identifies a new universality class distinct from the Barabási-Albert model, exhibiting non-scaling behavior in terms of $\bar{k}(t)$. The fact that $\bar{k}(t)$ follows a scaling pattern as a power of $\log(t)$ for the large enough times is similar to the \textit{fitness} model proposed by Bianconi and Barabási~\cite{bianconi2001competition}, although their degree distribution follows a stretched exponential. We observe from Fig.~\ref{fig:mean_field} a very good agreement for $\bar{k}$ between the $p$-CDA model, and the mean-field theory results. The inset reveals that $\xi_p$ is an increasing function of $p$.

\begin{figure}[t] 
  \centering
  \includegraphics{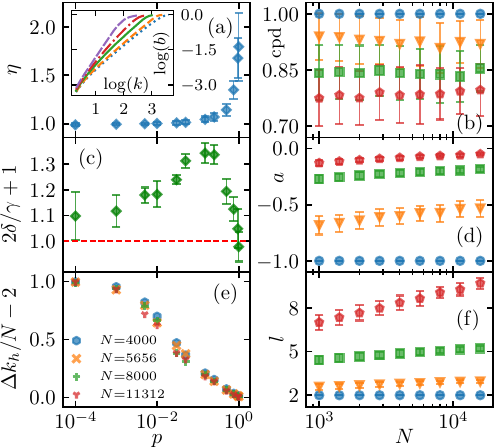}
  \caption{(a) The value of the exponent $\eta$ defined in $b(k) \sim k^{\eta}$ depends on $p$. Inset shows plots of this relation for the $p$ values of 0.0001, 0.15, 0.5, 0.75, 1 corresponding to the curves plotted from right to left, respectively. (b), (d), and (f) represent CPD, average shortest path length, and assortativity as functions of $N$ for different $p$ values: 0 (circles), 0.05 (triangles), 0.5 (squares), and 1 (pentagons). (c) A test for the Barthélemy's conjecture: the horizontal dashed line shows $\frac{2\delta}{\gamma+1}=1$. Except for 
$p=0.95$ and $p=1$, where the error bars include $1$, we observe that $\frac{2\delta}{\gamma+1} \geq 1$ for all other $p$ values. (e) Shows the degree difference between the two largest hubs of the system, divided by $N-2$, as a function of $p$.}
  \label{fig:globals}
\end{figure}

We also examined the correlation between degree and betweenness of the nodes, the analysis of which is presented in Fig.~\ref{fig:globals}, showing that $\eta$ decreases from $2$ to $1$ as we go from the Barabási-Albert limit to $p=0$. The \textit{central point dominance} (CPD) is expressed as~\cite{freeman1977set}
\begin{equation}
\text{CPD}(N)=\frac{1}{N-1}\sum_{i=1}^N\left(b_{\text{max}}-b_j\right),
\end{equation}
where $b_{\text{max}}$ represents the maximum betweenness in the network, corresponding to the betweenness of the super-hub for a network of size $N$. For the Barabási-Albert network CPD remains constant as the network expands, a feature that is observed in some scale free networks~\cite{adami2024dandelion}. Figure~\ref{fig:globals}(b) displays CPD as a function of $N$, which becomes a constant for large $N$ for all $p$ values. In Fig.~\ref{fig:globals}(c) we observe that Barthélemy's conjecture does not hold for large $p$ values, although the networks in the $p$-CDA model are trees. Our model also predicts higher values of disassortativity for small values of $p$. Specifically, the assortativity ($a$) of networks changes from $0$ to $-1$ as $p$ varies from $1$ to $0$, as shown in Fig.~\ref{fig:globals}(d). The degree difference between the two largest hubs of the network (the gap function $\Delta k_h$) is depicted in Fig.~\ref{fig:globals}(e) for various $N$ values in terms of $p$, exhibiting an exponential decay. We see that the gap closes as $p \to 1$ and is maximal ($=N-2$) when $p \to 0$. We also examined the average shortest path length, $l$, in the model concluding that lowering $p$, decreases $l$  since the super-hub connects almost all nodes. For intermediate $p$ values, the linear behavior of $l$ in terms of $N$ in semi-logarithmic scale in Fig.~\ref{fig:globals}(f)  reveals that $l\propto \ln N$ while for the considered system sizes it is a decreasing function of $p$. Note that for $p=1$, $l \propto \frac{\ln N}{\ln \ln N}$~\cite{fronczak2004average}.


\begin{figure}[t] 
	\centering
	\includegraphics{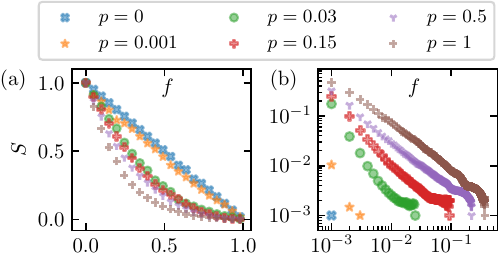}
	\caption{Sustainability $S$ for $N=1000$ in terms of $f$, the strength of attacks, for (a) random attacks and (b) targeted attacks (log-log scale).}
	\label{fig:attacks}
\end{figure}

In the context of network design, the $p$-CDA model it is of interest  addressing network vulnerability by investigating the network’s susceptibility/sustainability to different types of attacks. Depending on the nature of the attack, the likelihood of network failure varies relative to the Barabási-Albert model. In the case of targeted attack, nodes are removed based on their degree in the initial graph, starting with those having the highest degree, while in the random attack strategy, the nodes to be removed are chosen randomly. At the end, we analyze the sustainability, $S$, of networks via the size of the largest cluster in the network normalized by the size of the initial graph in terms of $f$ representing the number of attacks normalized by the total size of the network~\cite{albert2000error, Cohen2000, callaway2000network}. Figure~\ref{fig:attacks} presents results averaged over 100 realizations for each $p$ value. We observe that networks with smaller $p$ values are more resilient to random attacks, whereas networks with larger $p$ values show greater resilience to targeted attacks. The vulnerability of small-$p$ networks to targeted attacks can be mitigated by increasing the number of connections each new node creates upon addition.

\textbf{Concluding remarks:}
While the original goal of the Barabási-Albert model was to provide a mechanism for constructing scale-free complex networks using a preferential attachment scheme, it falls short of producing exponents that align with those observed in scale-free real-world networks, where $\gamma$ ranges from $1.5$ to $3$ and $\delta$ ranges from $1.37$ to $2$. This discrepancy indicates that scale-free complex networks cannot be constructed based solely on degree centrality.

In this letter, we discuss the impact of incorporating betweenness centrality directly into the growth equations by introducing a free external parameter $p$, which interpolates between the Barabási-Albert model ($p=1$) and a star graph ($p=0$). This marks a relevant shift in the Barabási-Albert universality class leading to a \textit{stars-with-filament} structure. In mean-field the $p$-CDA model, is characterized by a power-law dependence on $\log t$ for $\bar{k}$ (according to Eq.~\ref{Eq:averageK}) 
and a logarithmic dependence of $l$ on $N$. This mean-field theory successfully matches the numerical results (see Fig.~\ref{fig:mean_field}). Remarkably, this model offers the possibility to describe the varying exponents observed in most scale-free networks. This model not only allows for a continuous exploration of various universality classes of scale-free real networks but also provides insights into their topology and structures. We present strong numerical evidence that the $p$-CDA model is sufficiently accurate to effectively describe these real-world networks by adjusting $p$. Figure~\ref{fig:phase_space} shows the effective $p$ values, an assessment of Barthélemy's conjecture, and a comparison with the $p$-CDA model for 47 scale-free real-world networks analyzed here. The structural variations of the networks were assessed by examining  CPD, assortativity, average shortest path length, and the degree difference between the two largest hubs in the network, as illustrated in Fig.~\ref{fig:globals}. 

Here, we discuss the possible application of our model to some scale-free real networks. Our analysis of the Person-Country Affiliations network~\cite{auer2007dbpedia} reveals an optimal $p$ value of $0.1$ (with the exponents of $\gamma=1.49\pm0.03$, $\delta=1.85\pm0.02$, and $\eta=1.05\pm0.04$), showing the importance of the betweenness relative to that of degree. This is because individuals with high betweenness centrality in this network serve as key intermediaries between different countries, facilitating collaborations, exchanges, or negotiations. Their position allows them to bridge gaps and connect disparate groups, thereby enhancing their influence on international relations. Connecting to such influential individuals increases the visibility of new nodes within the network, which explains why the effective $p$ is small in this network.

Another example is the bipartite user-page network extracted from Wiktionary, considering the word ``or"~\cite{wikimedia_dumps}, from the three languages: French, German (if available), and English. A gap between the degree of the super-hub and the cut-off degree is observed in this network, which has already been predicted by the $p$-CDA model for scale-free networks with small $p$ values (Fig.\ref{fig:globals}e). In this network, a user is linked to a page if he/she makes edits to it. We evaluate the exponents $\gamma = 1.5 \pm 0.1$, $\delta = 1.53 \pm 0.04$, and $\eta = 1.06 \pm 0.02$ for this network, corresponding to a $p$ value of $0.05$. Given the limited number of pages available for editing, it is intuitive to expect the presence of a super-hub with high betweenness centrality, meaning that some pages are edited more frequently by users. Such pages act as access points for users, indicating that they are popular or essential resources for understanding the term ``or" and thus attract more user edits. A similar description applies to the Internet~\cite{karrer2014percolation} and the Anybeat online social network~\cite{fire2013link}, which are consistent with an optimal $p$ value of $0.38$.

Two examples of the opposite case (high $p$) are the email networks from the Enron corpus~\cite{klimt2004enron} and a European research institution~\cite{leskovec2007graph}, for which the optimal $p$ values are $0.82$ and $0.58$, respectively, indicating that degree centrality plays a more vital role. In the Enron email network, nodes represent email addresses, and node $i$ is connected to node $j$ if $i$ has sent at least one email to address $j$. In this context, degree centrality indicates how many direct connections (emails sent and received) a particular node (email address) has. A node with high degree centrality has sent (or received) emails to (or from) many different addresses, suggesting it is well-connected. Therefore, a new node is more likely to receive an email from a high-degree node. The same reasoning applies to the European research institution network.

\bibliography{refs}
\appendix

\section{$q$-moments analysis in complex networks}
In this section, we present an analysis on the $q$-moment of the degree and the betweenness centralities for the scale free (SF) networks defined as follows:
\begin{equation}
Z_{q}(t)\equiv \frac{1}{2t}\sum_{i=1}^tk_i^{q},\ \mathcal{Z}_{q}(t)\equiv t^{q-1}\sum_{i=1}^tb_i^{q}
\end{equation}
where, $t$ represents time and $q$ is an external parameter that plays the role of the moment. Defining $t \equiv N-1$, where $N$ is the network size we have the following asymptotic limits:
\begin{equation}
\begin{split}
&2\lim_{q\to 0}Z_q(t)=\lim_{q\to 0}\mathcal{Z}_q(t)=1, \ \ \lim_{q\to 1}Z_q(t)=1, \\
&\lim_{q\to \infty}Z_q(t)=\frac{1}{2}P(k_{\text{max}},t)\left(\frac{k_{\text{max}}}{t}\right)^q.
\end{split}
\label{Eq:limits}
\end{equation}
where, $k_{\text{max}}$ is the maximum $k$ value, and $P(k,t)$ is the normalized distribution of degrees at time $t$ (the same is defined for the betweenness $P(b,t)$). In large $t$ values one can calculate the average $q$-moments as follows:
\begin{equation}
\begin{split}
\bar{Z}_q(t)=\frac{1}{2t}\sum_k P(k,t)k^{q}\Delta k & \to \frac{1}{2t}\int_1^{k_{\text{max}}}p(k,t)k^qdk,\\
\bar{\mathcal{Z}}_q(t)=t^{q-1}\sum_k P(b,t)b^{q}\Delta b & \to t^{q-1}\int_{b_{\text{min}}}^{b_{\text{max}}}p(b,t)b^qdb,
\end{split}
\end{equation}
where, in the final step, we used the probability distribution function defined as $P(x,t)\Delta x \to p(x,t)dx$ ($x = k, b$), with $\Delta k = 1$. \\

An instructive example is the SF networks, for which the distribution follows a power law: $p(k,t) = A_k(t) k^{-\gamma}$ and $p(b,t) = A_b(t) b^{-\delta}$, where $A_x(t)$ ($x = k, b$) are time-dependent proportionality parameters. In this case, we have
\begin{equation}
\bar{Z}_q(t)=\frac{A_k(t)}{2t} \int_1^{k_{\text{max}}}k^{q-\gamma}dk=\frac{A_k(t)\left(k_{\text{max}}^{1+q-\gamma}-1\right)}{2t(1+q-\gamma)},
\label{Eq:Z1}
\end{equation}
and also
\begin{equation}
\bar{\mathcal{Z}}_q(t)=t^{q-1}A_b(t) \int_{b_{\text{min}}}^{b_{\text{max}}}b^{q-\delta}db=\frac{A_b(t)\left(b_{\text{max}}^{1+q-\delta}-b_{\text{min}}^{1+q-\delta}\right)}{t^{1-q}(1+q-\delta)},
\label{Eq:Z2}
\end{equation}
Using the limits checked in the Eq.~(\ref{Eq:limits}) one finds:
\begin{equation}
\begin{split}
&\frac{A_k(t)}{\gamma-1}\left(1-k_{\text{max}}^{1-\gamma}\right)=t,\\
&\frac{A_k(t)}{2(2-\gamma)}\left(k_{\text{max}}^{2-\gamma}-1\right)=t.
\end{split}
\end{equation}
giving
\begin{equation}
\begin{split}
\frac{1-k_{\text{max}}^{1-\gamma}}{k_{\text{max}}^{2-\gamma}-1}=\frac{\gamma-1}{2(2-\gamma)}\\
A_k(t)=\left(\frac{\gamma-1}{1-k_{\text{max}}^{1-\gamma}}\right)t, 
\end{split}
\end{equation}
From which, one realizes that $A_k(t)$ is proportional to time, and $Z_q(t)$ is independent of time, as seen from Eq.~(\ref{Eq:Z1}). The same calculations hold true for $\mathcal{Z}_q(t)$ as well. \\

We also have
\begin{equation}
A_b(t)=\left(\frac{\delta-1}{b_{\text{min}}^{1-\delta}-b_{\text{max}}^{1-\delta}}\right)t.
\end{equation}
Using the scaling relation between $k$ and $b$, i.e.
\begin{equation}
b=A_{bk}t^{-\tau}k^{\eta} 
\end{equation}
where $A_{bk}$ is a constant ($\delta>1$, $\tau$ is an exponent, and $k_{\text{min}}\equiv 1$)
\begin{equation}
A_b(t)=\left(\frac{\delta-1}{A_{bk}^{1-\delta}(1-k_{\text{max}}^{\eta(1-\delta)})}\right)t^{1+\tau(1-\delta)}.
\end{equation}
Incorporating this into Eq.~(\ref{Eq:Z2}), one finds $\mathcal{Z}_q(t)$. More precisely
\begin{equation}
\begin{split}
& \bar{Z}_q(t)\equiv \bar{Z}_q=\frac{(\gamma-1)\left(k_{\text{max}}^{1+q-\gamma}-1\right)}{2(1-k_{\text{max}}^{1-\gamma})(1+q-\gamma)}\\
&\bar{\mathcal{Z}}_q(t)\equiv \bar{\mathcal{Z}}_q=\frac{(\delta-1)A_{bk}^{(q)}\left(k_{\text{max}}^{\eta(1+q-\delta)}-1\right)}{(1-k_{\text{max}}^{\eta(1-\delta)})(1+q-\delta)}t^{q(1-\tau)}.
\label{Eq:MFrelations}
\end{split}
\end{equation}
For most SF networks (including the BA CN), $\tau = 1$, which results in the time-independence of $\mathcal{Z}_q(t)$. These formulas can be used to determine the moments of SF CNs, and we also use them in our mean-field arguments. Before ending this section, we mention that for a pure SF network, The same integration progress as above gives the average degree of nodes $\bar{k} $ (first moment of $k$) as a function of $N$ and $\gamma$ in the limit $N\to \infty$ reads (note that the distribution is Gaussian for $\gamma> 3$)
	\begin{equation}
		\bar{k} =\left\lbrace \begin{matrix}
			AN^{\alpha}\left(1-\gamma\right)/\left(2-\gamma\right) & \gamma<1,\\
			AN^{\alpha}/\left(\alpha\ln N+\xi\right) & \gamma=1,\\
			\left(AN^{\alpha}\right)^{2-\gamma}\left(\gamma-1\right)\left(2-\gamma\right) & 1<\gamma<2,\\
			\alpha\ln N+\xi & \gamma=2,\\
			(\gamma-1)/(\gamma-2) & 2<\gamma\le3,
		\end{matrix}\right.
	\end{equation}
where we have used the fact that the maximal node degree $k_{\text{max}}$ scales with the network size as $k_{\text{max}}=AN^{\alpha}$, defining the exponent $\alpha$, and $\xi\equiv \ln A$.
\section{mean-field theory for $p$-CDA}\label{sec:mean_filed_appendix}
\begin{figure*}[t] 
  \centering
  \includegraphics{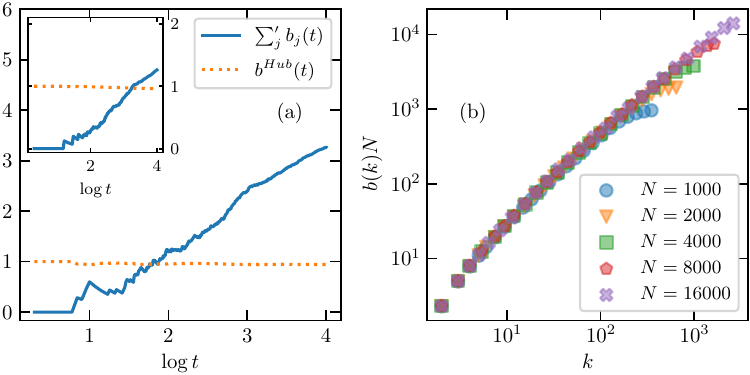}
  \caption{(a) Time evolution of the BC of hub and the cumulative residue BC for $p=0.5$ (main) and $p=0.1$ (inset) in terms of $\log t$. (b) A data collapse analysis for BC-DC correlation relation for $p=0.5$ in log-log scale. A similar scaling relation holds for other $p$ values.}
  \label{fig:mean_field_BC}
\end{figure*}
Now, we consider the mean-field (MF) theory for the $p$-CDA model. The growth of the $i$th on the MF level is governed by the following expression
\begin{equation}
\frac{dk_i}{dt}=p\Pi(k_i,t)+(1-p)\Pi(b_i,t),
\end{equation}
where $\Pi(x_i,t)\equiv \frac{x_i}{\sum_jx_j}$, $x_i=k_i, b_i$ are the degree and the betweenness of the $i$th node, respectively. This equation states that the growth of $k_i$ is determined by $\Pi(k_i,t)$ with probability $p$, and by $\Pi(b_i,t)$ with probability $1-p$. Using the $q$-variances defined in the previous section, this equation can also be written as
\begin{equation}
\frac{dk_i}{dt}=\frac{p}{2Z_{q=1}(t)}\frac{k_i}{t}+(1-p)\frac{b_i}{\mathcal{Z}_{q=1}(t)}.
\label{Eq:diffEq}
\end{equation}
Before analyzing the solutions of this equation, it is worth mentioning that for SF networks:
\begin{equation}
b_j=A_{bk}^{(p)} t^{-\tau_p}k_j^{\eta_p},
\label{Eq:b-kCorrelation}
\end{equation}
where $A_{bk}^{(p)}$ is a proportionality constant introduced in the previous section, and $\tau_p$ and $\eta_p$ are some exponents. To capture the dynamics of the hub node and the rest nodes separately, we decompose the summation over the BCs as follows:
\begin{equation}
\sum_j b_j(t)=b^{\text{Hub}}_p(t)+\sum'_{j}b_{j}(t),
\end{equation}
where $b^{\text{Hub}}_p$ is the betweenness of the hub node, and $\sum'$ denotes summation over all nodes except the hub, called the \textit{residual betweenness}. Our numerical results show that for sufficiently large times
\begin{equation}
\begin{split}
& b^{\text{Hub}}_p(t)\to \bar{b}^{\text{Hub}}_p=\text{constant}\\
&\sum'_{j}b_{j}(t)\to C_p+B_p\log t,
\end{split}
\end{equation}
where $B_p$, $C_p$, and $\bar{b}^{\text{Hub}}_p$ are constants. In Figs.~\ref{fig:mean_field_BC}(a) and~\ref{fig:mean_field_BC}(b), we show these functions, which match the expected trends—i.e., constant for the hub betweenness and logarithmic growth for the residual accumulated betweenness. Figure~~\ref{fig:mean_field_BC}(b) confirms the scaling relation in Eq.~(\ref{Eq:b-kCorrelation}) with $\tau_p = 1$ (for all $p$ values) using the data collapse technique.\\

Incorporating these equations into Eq.~(\ref{Eq:diffEq}), along with the Eq.~(\ref{Eq:b-kCorrelation}), we obtain
\begin{equation}
\frac{dk_i}{dt}=\frac{p}{2Z_{q=1}(t)}\frac{k_i}{t}+(1-p)\frac{A_{kb}k_i^{\eta_p}}{t(\bar{b}_p^{\text{Hub}}+C_p+B_p\log t)}.
\end{equation}
In the $t\to \infty$ limit the logarithmic term overcomes the other two terms in the denominator, and given that $Z_{q=1}(t)=1$ one finds
\begin{equation}
\text{d}\log k_i=\frac{p}{2}\text{d}\log t+(1-p)\frac{A_{bk}k_i^{\eta_p-1}}{B_p}\text{d}(\log \log t).
\end{equation}
or equivalently
\begin{equation}
\frac{\text{d}Q_i}{\text{d}T}=\frac{p}{2}+D_p\frac{e^{(\eta_p-1)Q_i}}{T},
\end{equation}
where $D_p\equiv \frac{(1-p)A_{bk}}{B_p}$, $Q_i\equiv \log k_i$, and $T\equiv \log t$. The solution of the above equation is
\begin{equation}
Q_i(t)=\frac{pT}{2}-\frac{\log \left(\frac{p}{2} D_p (\eta_p-1)^2 \left(\frac{\text{Ei}(\frac{p}{2} T (\eta_p-1) )}{-\frac{p}{2} (\eta_p-1) }-c_1\right)\right)}{\eta_p-1 },
\end{equation}
where $c_1$ is a constant, and $E_i(z)$ is an exponential integral function~\cite{gradshteyn2014table}. We see that this solution is very complicated. Fortunately, for the intermediate times, we can use another approximation which makes the solution much easier.

\subsection{An approximation for intermediate times}
As mentioned in the previous subsection, the residual accumulated betweenness changes smoothly over time. For intermediate times (up to $\sim 10^4$), the logarithmic term grows only up to certain limited numbers. To enable more practical calculations, we approximate the residual betweenness term by its average over the time interval, represented by $\bar{b}_{r}$. This approximation reads
\begin{equation}
\begin{split}
  \mathcal{Z}_{q=1}(t) & \approx \frac{1}{\Delta t}\int_{t-\frac{\Delta t}{2}}^{t+\frac{\Delta t}{2}}\mathcal{Z}_{q=1}(t')dt'
  \end{split} 
\end{equation}
where $\Delta t$ is the interval of interest. This is equivalent to:
\begin{equation}
\begin{split}
  \sum_j b_j(t)&\approx \frac{1}{\Delta t}\int_{t-\frac{\Delta t}{2}}^{t+\frac{\Delta t}{2}}\left(b^{\text{Hub}}_p(t')+\sum'_j b_j(t')\right)dt'\\
  &= \bar{b}^{\text{Hub}}_p+\bar{b}_{r}.  
\end{split} 
\end{equation}
Substituting this relation into Eq.~(\ref{Eq:diffEq})
implies
\begin{equation}
\frac{dk_i}{dt}=p\frac{k_i}{2t}+\frac{A_{kb}^{(p)}(1-p)}{\bar{b}^{\text{Hub}}_p+\bar{b}_r}\frac{k_i^{\eta_p}}{t^{\tau_p}}.
\end{equation}

As we have already mentioned, our numerical simulations reveal that $\tau_p = 1$ for all $p$ values. Thus, by replacing $k_i$ with its average $\bar{k}$, we obtain
\begin{equation}
\frac{d\bar{k}}{dt}=\frac{c_p^{(1)}\bar{k}+c_p^{(2)}\bar{k}^{\eta_p}}{t},
\label{Eq:diff}
\end{equation}
where $c_p^{(1)} \equiv \frac{p}{2}$ and $c_p^{(2)} \equiv \frac{A_{kb}^{(p)}(1-p)}{\bar{b}^{\text{Hub}}_p + \bar{b}_r}$. The solution is obtained through a simple integration:
\begin{equation}
\int_{k_0}^{k}\frac{d\bar{k}}{\frac{c_p^{(1)}}{c_p^{(2)}}\bar{k}+\bar{k}^{\eta_p}}=c_p^{(2)}\log \frac{t}{t_0}
\label{eqt:mean_field}
\end{equation}
where $k_0$ and $t_0$ are some initial values. This integral can be done using the following identity:
\begin{equation}
    \int \frac{dx}{ax+x^b} = \frac{1}{a(1-b)}\ln \left[1+ax^{1-b}\right],
\end{equation}
resulting to the following relation
\begin{equation}
 \ln \frac{1+\frac{c_p^{(1)}}{c_p^{(2)}}k^{1-\eta_p}}{1+\frac{c_p^{(1)}}{c_p^{(2)}}k_0^{1-\eta_p}}   = c_p^{(1)}(1-\eta_p)\ln \frac{t}{t_0}.
\end{equation}
Looking at the solution in some asymptotic limits helps much. When $p$ tends to 1 (corresponding to the BA model), $c_p^{(2)}$ tends to zero and the second term in the right hand side of Eq.~\ref{Eq:diff} (the term proportional to $k^{\eta}$) vanishes. As a result, one can recover the MF equation of the BA model, with $k\propto t^{\frac{1}{2}}$. For other values of $p$, the fact that $\eta_p > 1$ for all $p$ values implies that the term $\bar{k}^{\eta_p}$ is dominant for sufficiently large $k$ values. Therefore, by ignoring the first term in the denominator of the integrand on the left-hand side, one finds, to leading order,
\begin{equation}
\frac{1}{\eta_p-1}\left[\frac{1}{\bar{k}_0^{\eta_p-1}}-\frac{1}{\bar{k}^{\eta_p-1}}\right]=c_p^{(2)}\log \frac{t}{t_0}.
\end{equation}
This finally leads to
\begin{equation}
\bar{k}=\left(a_p-b_p\log t\right)^{-\xi_p},
\end{equation}
where $a_p\equiv \frac{1}{\bar{k}_0^{\eta_p-1}}+(\eta_p-1)c_p^{(2)}\log t_0$, $b_p\equiv (\eta_p-1)c_p^{(2)}$, and $\xi_p\equiv \frac{1}{\eta_p-1}$. Since these parameters are complicated functions of $t_0$, $k_0$ $p$, $\eta_p$, $A_p$, and $\bar{b}_p^{\text{Hub}}$, we consider them as fitting functions. This relation is our main finding to be used in the paper.

\subsection{Average degree calculation for SF networks without a hub}
In this section, we ignore the special role of the hub. In this case we have
\begin{equation}
\begin{split}
\mathcal{Z}_q &=t^{q-1}\sum_{i=1}^t(A_{bk}^{(p)}t^{-\tau_p}k^{\eta_p})^q\\
&=\frac{1}{t}\left(\frac{tA_{bk}^{(p)}}{t^{\tau_p}}\right)^{q}\sum_{i=1}^tk^{\eta_pq}=2\left(A_{bk}^{(p)}\right)^qZ_{q\eta_p}(t)
\end{split}
\end{equation}
where in the last line we have used the fact that $\tau_p=1$. This leads us to write the equations as follows:
\begin{equation}
\frac{dk_i}{dt}=p\frac{k_i}{2t}+(1-p)\frac{k_i^{\eta_p}}{2t^{\eta_p}Z_{\eta_p}(t)}.
\end{equation}
One can use   Eq.~\ref{Eq:MFrelations} (which is independent of time) to find
\begin{equation}
\frac{dk_i}{dt}=p\frac{k_i}{2t}\left(1+\frac{\epsilon_p}{\bar{Z}_{\eta}}\frac{k_i^{\eta_p-1}}{t^{\eta_p-1}}\right),
\label{Eq:MF2}
\end{equation}
where $\epsilon_p\equiv \frac{1-p}{p}$. Note that for $t\to\infty$, the second term in the parentheses becomes a sub-leading term (since $\eta_p>1$), so that
\begin{equation}
\bar{k}\propto t^{\beta_p},
\end{equation}
where $\beta_p\equiv \frac{p}{2}$, which is again consistent with the well-known exponent $\lim_{p\to 1}\beta_p=\frac{1}{2}$ in the BA limit~\cite{BARABASI1999173}.\\

Many other functions can be calculated using the $q$-moment functions, like the  $q$-Renyi entropy. It is given as
\begin{equation}
\mathcal{R}e_q(t)=\frac{1}{1-q}\ln \left(2tZ_q(t)\right),
\end{equation}
from which one can calculate the Shannon entropy. Using Eq.~\ref{Eq:MFrelations}, one finds
\begin{equation}
\mathcal{R}e_q(t)=f_1(q)+f_2(q)\ln t,
\end{equation}
where $f_1(q)\equiv\frac{1}{1-q}\left[\ln(2\bar{Z}_q)\right]$, and $f_2(t)\equiv \frac{1}{1-q}$. We see that it grows logarithmically with time. One may consider non-overlapping intervals with Dijkstra radius $\delta$ and define, for each segment $I$,
\begin{equation}
\mu_I\equiv \frac{\sum_{i\in I}k_i}{t},
\end{equation}
where $t$ is the total system size. The generalized $q$-moment function is then defined as: 
\begin{equation}
Z_q(\delta,t)=\sum_I\left(\mu_I\right)^q.
\end{equation}
In the case of a SF system, this function may behave as:
\begin{equation}
Z_q(\delta,t)\propto \delta^{\alpha_q}, \ \text{so that:}\ \alpha_q=\lim_{\delta\to 0} \frac{\log Z_q(\delta,t)}{\log \delta}.
\end{equation}
Importantly, the generalized $q$-dimension is defined as:
\begin{equation}
D_q\equiv \frac{\alpha_q}{q-1},
\end{equation}
such that the fractal dimension of the network is given by $D_f=\lim_{q\to 0}D_q$.

\section{Analysing of Real networks}
In this section, we examine 47 real networks previously utilized in our main paper to assess the validity of our proposed model. The data for these networks was sourced from Netzschleuder~\cite{tiago_p_peixoto_2023_7839981}, a catalog and repository of network datasets, and statistically analyzed using graph-tool~\cite{peixoto_graphtool_2014}. This repository includes 286 datasets, each containing at least one network. The selection process for the 47 networks from the mentioned repository followed a few criteria: (I) the number of nodes had to be sufficiently large (approximately $\gtrapprox 10^4$) to ensure smoother distributions with minimal noise; (II) the networks analyzed needed to exhibit scale-free, power-law behavior; and (III) the calculated exponents of these scale-free networks had to fall within the range of exponents derived from the $p$-CDA model, allowing for a meaningful comparison between the $p$-CDA model and the real networks.\\

After finding candidate networks, we derive three exponents, $\gamma$ and $\delta$ as well as $\eta$ from the degree and betweenness distributions and their correlation. Subsequently, we assess the suitability of different values of $p$ for these networks based on the model we have constructed. To do so, we introduce the equation 
\begin{equation}
    f_{\zeta}(p) = \left(\zeta_{\text{model}}(p)-\zeta_{\text{rn}}\right)^2
    \label{eqt:function}
\end{equation}
which measures the difference between the modeled exponent ($\zeta_{\text{model}}(p)$) and the actual exponent obtained from the real networks ($\zeta_{\text{rn}}$) squared. By calculating this function across various $p$ values (13 different numbers within the range of zero and one we have opted in our numerical calculations), we aim to pinpoint the specific value at which a minimum occurs. This minimum corresponds to the optimal value of $p$ for the exponent $\zeta$, which may correspond to $\gamma$, $\delta$ or $\eta$. The $\zeta_{\text{model}}$ represents the calculated exponent from our model, while $\zeta_{\text{rn}}$ signifies the exponent observed in the real networks.\\ 

Calculated exponents and their corresponding 
$p$-values, obtained from Eq.~(\ref{eqt:function}), are reported in  Table~\ref{tab:network_comparison}. All 47 networks in this table are indexed with numbers from 1 to 47. These indices allow for easy identification of the networks in the $\delta-\gamma$ phase-space in the Fig. 3 of the main paper. We also calculated the relations $\frac{2\delta}{\gamma+1}$ (derived from Barthelemy's conjecture) and $\frac{\eta(\delta-1)}{(\gamma-1)}$, which are expected to equal one for scale-free tree networks.\\

\begin{figure*}[t] 
  \centering
  \includegraphics{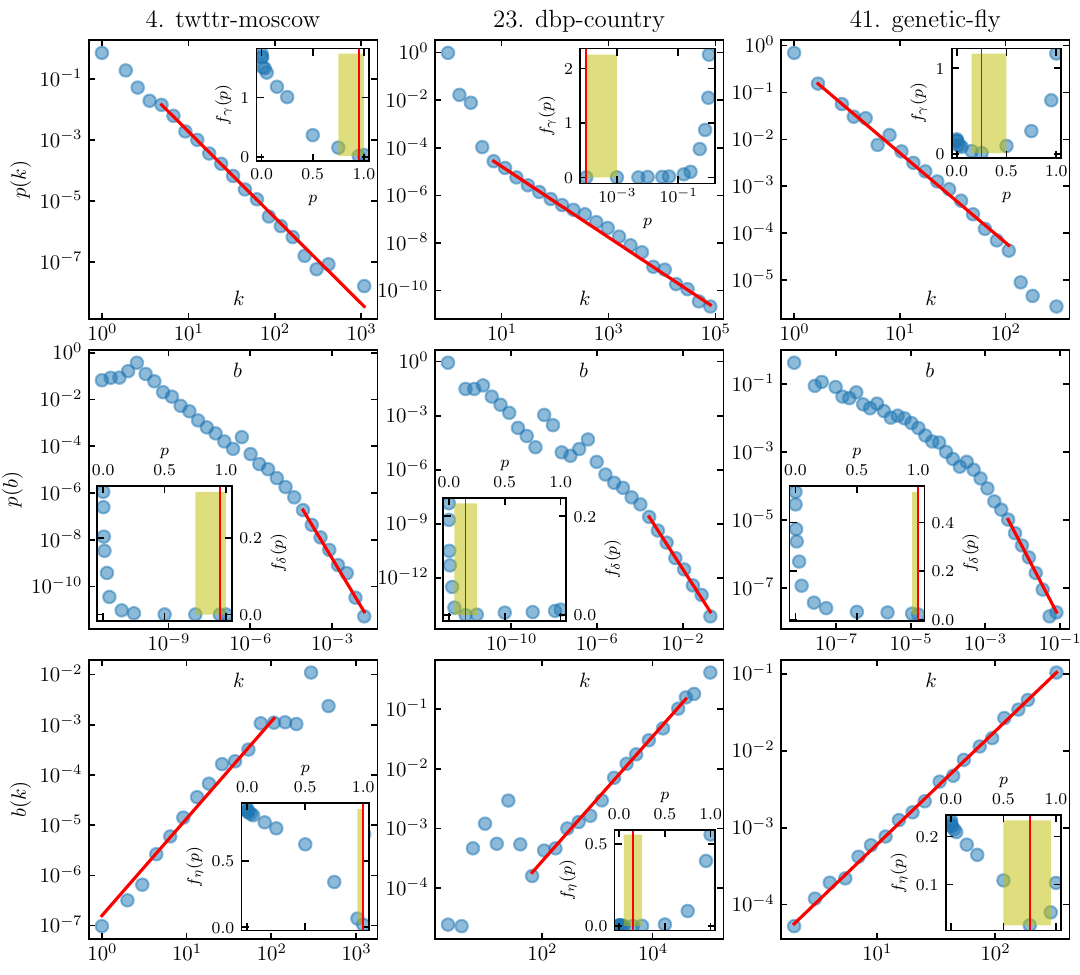}
  \caption{The DC and BC distributions, along with their correlations for three real networks, are represented as follows: ``Twitter, MoscowAthletics2013"~\cite{omodei2015characterizing} (left column), ``Person-Country Affiliations"~\cite{auer2007dbpedia} (middle column), and ``genetic-fly — MIST protein interaction"~\cite{hu2018molecular} (right column). Based on the calculated exponents derived from the scaling functions $p(k)\sim k^{-\gamma}$, $p(b)\sim b^{-\delta}$, and $b(k)\sim k^{\eta}$, with the best fits shown by the solid lines, the corresponding $p$ values for the related scaling functions of each network can be determined using Eq.~(\ref{eqt:function}). The effective $p$ values obtained are 0.97 for the ``left", 0.10 for the ``middle", and 0.67 for the ``right" networks. The optimal $p$ values are indicated in the insets by vertical solid lines.}
  \label{fig:sample_RNs}
\end{figure*}

\begin{figure*}[t] 
  \centering
  \includegraphics{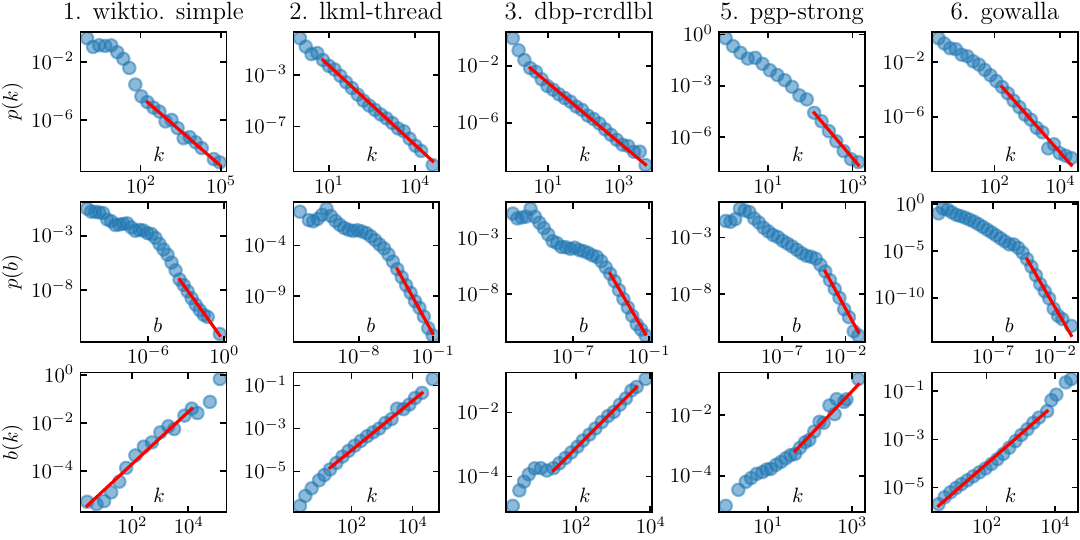}
  \caption{The DC and BC distributions, along with their correlations for five real networks, are shown across five columns. From left to right, the columns represent the following networks: ``wiktio. simple" (1st column), ``kml-thread" (2nd column), ``dbp-rcrdlbl" (3rd column), ``pgp-strong" (4th column), and ``gowalla" (5th column). The solid lines represent the best fits used to determine the scaling exponents $\gamma$, $\delta$, and $\eta$.}
  \label{fig:real_networks_1}
\end{figure*}

\begin{figure*}[t] 
  \centering
  \includegraphics{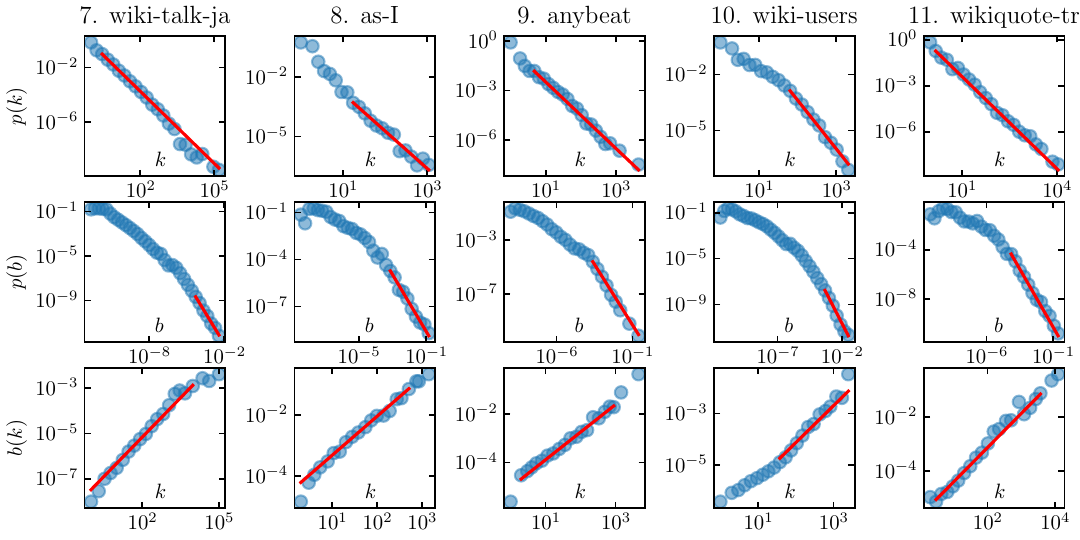}
  \caption{The DC and BC distributions, along with their correlations for five real networks, are shown across five columns. From left to right, the columns represent the following networks: ``wiki-talk-ja" (1st column), ``as-I" (2nd column), ``anybeat" (3rd column), ``wiki-users" (4th column), and ``wikiquote-tr" (5th column). The solid lines represent the best fits used to determine the scaling exponents $\gamma$, $\delta$, and $\eta$.}
  \label{fig:real_networks_2}
\end{figure*}

\begin{figure*}[t] 
  \centering
  \includegraphics{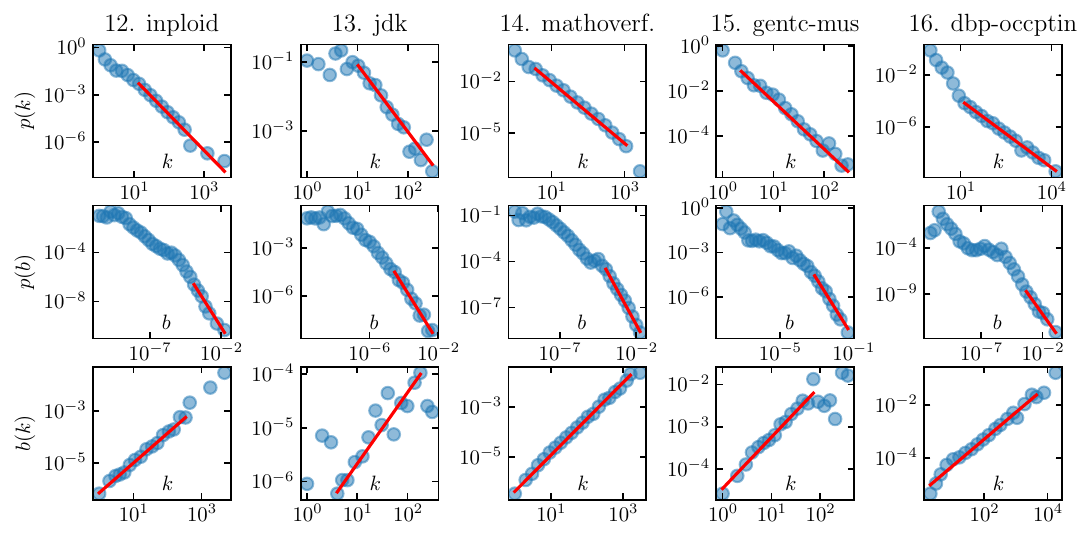}
  \caption{The DC and BC distributions, along with their correlations for five real networks, are shown across five columns. From left to right, the columns represent the following networks: ``inploid" (1st column), ``jdk" (2nd column), ``mathoverf." (3rd column), ``gentc-mus" (4th column), and ``dbp-occptin" (5th column). The solid lines represent the best fits used to determine the scaling exponents $\gamma$, $\delta$, and $\eta$.}
  \label{fig:real_networks_3}
\end{figure*}

\begin{figure*}[t] 
  \centering
  \includegraphics{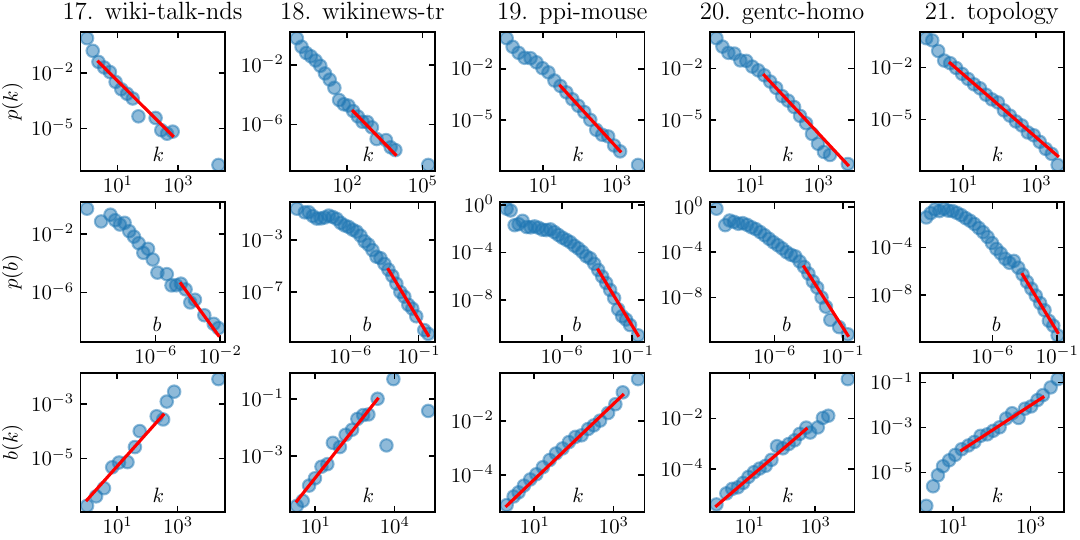}
  \caption{The DC and BC distributions, along with their correlations for five real networks, are shown across five columns. From left to right, the columns represent the following networks: ``wiki-talk-nds" (1st column), ``wikinews-tr" (2nd column), ``ppi-mouse" (3rd column), ``gentc-homo" (4th column), and ``topology" (5th column). The solid lines represent the best fits used to determine the scaling exponents $\gamma$, $\delta$, and $\eta$.}
  \label{fig:real_networks_4}
\end{figure*}

\begin{figure*}[t] 
  \centering
  \includegraphics{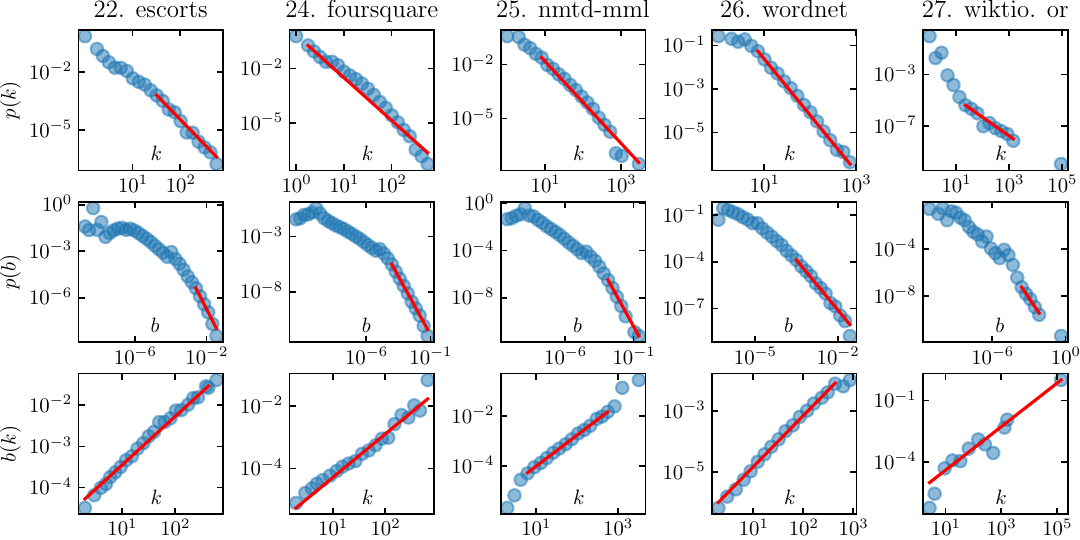}
  \caption{The DC and BC distributions, along with their correlations for five real networks, are shown across five columns. From left to right, the columns represent the following networks: ``escorts" (1st column), ``foursquare" (2nd column), ``nmtd-mml" (3rd column), ``wordnet" (4th column), and ``wiktio. or" (5th column). The solid lines represent the best fits used to determine the scaling exponents $\gamma$, $\delta$, and $\eta$.}
  \label{fig:real_networks_5}
\end{figure*}

\begin{figure*}[t] 
  \centering
  \includegraphics{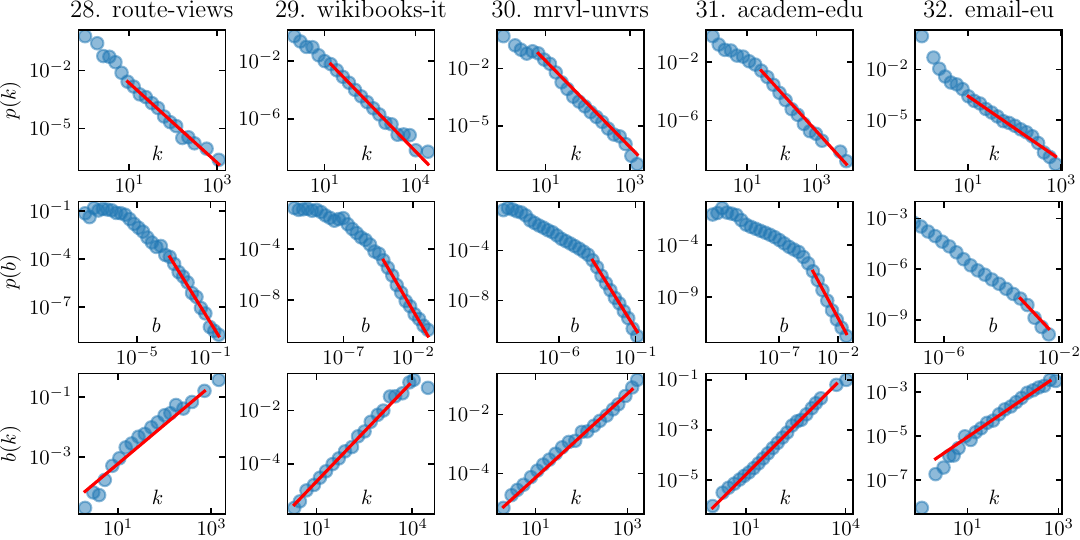}
  \caption{The DC and BC distributions, along with their correlations for five real networks, are shown across five columns. From left to right, the columns represent the following networks: ``route-views" (1st column), ``wikibooks-it" (2nd column), ``mrvl-unvrs" (3rd column), ``academ-edu" (4th column), and ``email-eu" (5th column). The solid lines represent the best fits used to determine the scaling exponents $\gamma$, $\delta$, and $\eta$.}
  \label{fig:real_networks_6}
\end{figure*}

\begin{figure*}[t] 
  \centering
  \includegraphics{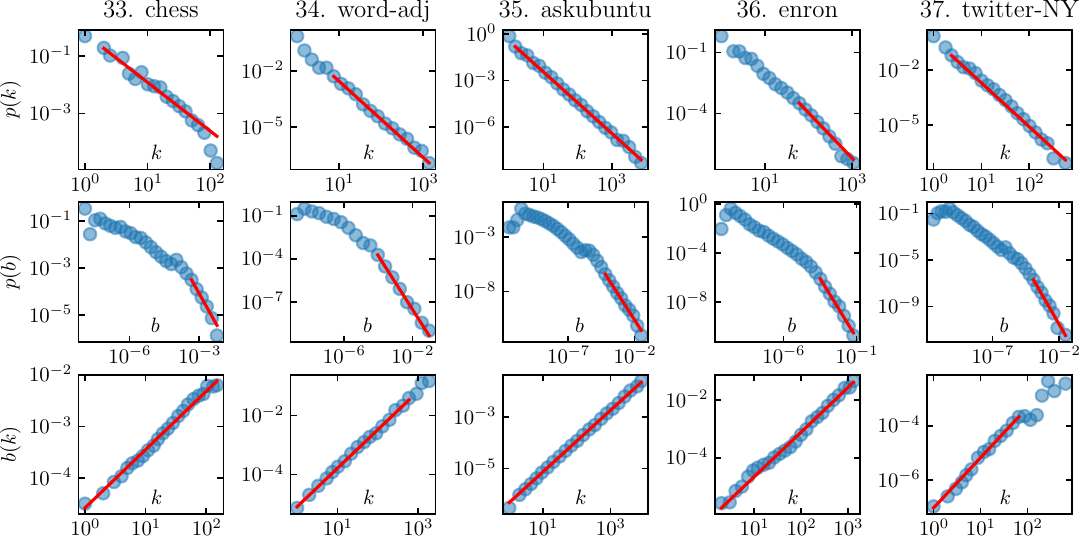}
  \caption{The DC and BC distributions, along with their correlations for five real networks, are shown across five columns. From left to right, the columns represent the following networks: ``chess" (1st column), ``word-adj" (2nd column), ``askubuntu" (3rd column), ``enron" (4th column), and ``twitter-NY" (5th column). The solid lines represent the best fits used to determine the scaling exponents $\gamma$, $\delta$, and $\eta$.}
  \label{fig:real_networks_7}
\end{figure*}

\begin{figure*}[t] 
  \centering
  \includegraphics{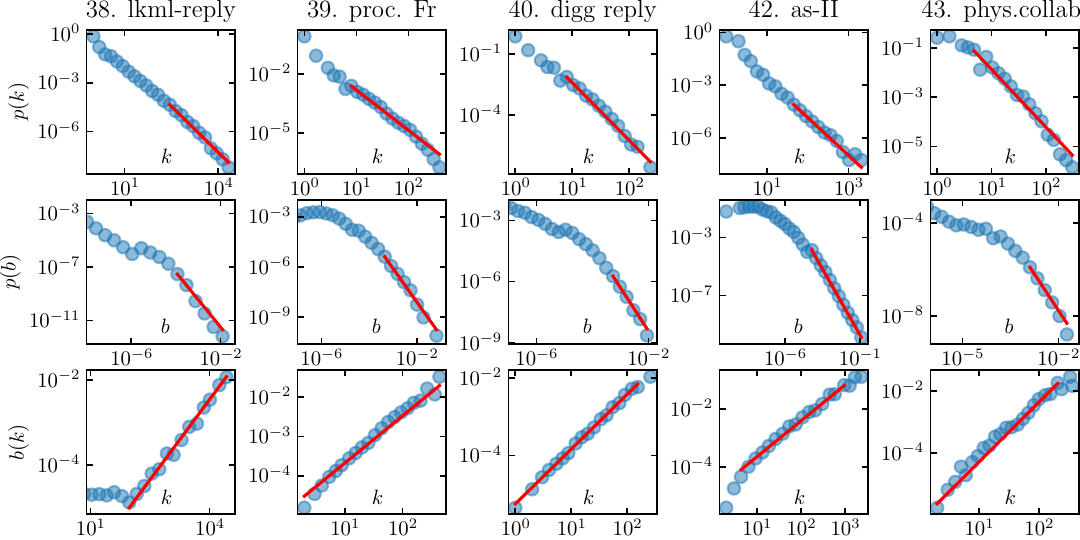}
  \caption{The DC and BC distributions, along with their correlations for five real networks, are shown across five columns. From left to right, the columns represent the following networks: ``lkml-reply" (1st column), ``proc. Fr" (2nd column), ``digg reply" (3rd column), ``as-II" (4th column), and ``phys.collab" (5th column). The solid lines represent the best fits used to determine the scaling exponents $\gamma$, $\delta$, and $\eta$.}
  \label{fig:real_networks_8}
\end{figure*}

\begin{figure*}[t] 
  \centering
  \includegraphics{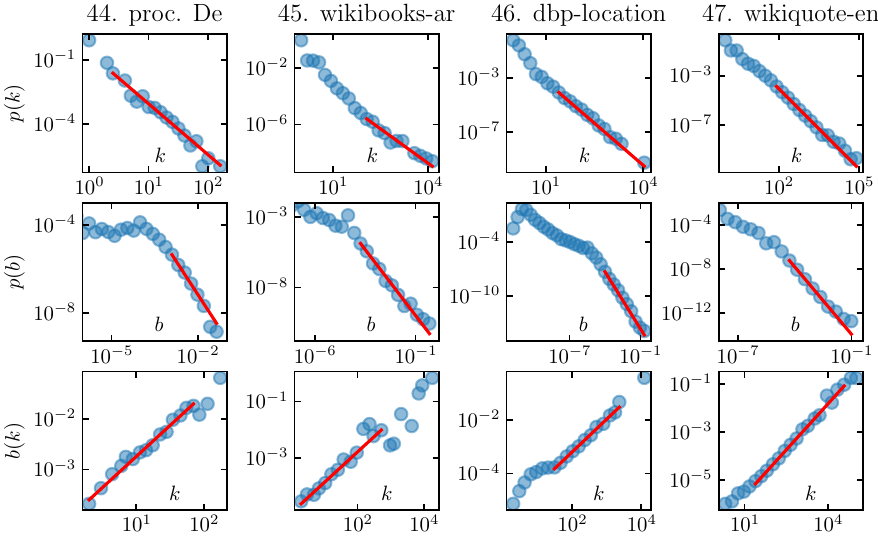}
  \caption{The DC and BC distributions, along with their correlations for four real networks, are shown across four columns. From left to right, the columns represent the following networks: ``proc. De" (1st column), ``wikibooks-ar" (2nd column), ``dbp-location" (3rd column), and ``wikiquote-en" (4th column). The solid lines represent the best fits used to determine the scaling exponents $\gamma$, $\delta$, and $\eta$.}
  \label{fig:real_networks_9}
\end{figure*}

Figure~\ref{fig:sample_RNs} presents the degree and betweenness distributions, along with their correlation, for three representative real networks. These networks exhibit scale-free properties, as evidenced by their power-law behavior. The solid lines in the main plots represent the best fits, with a minimum R-squared value of 0.95. After determining the corresponding exponents through the fitting process, we compute Eq.~(\ref{eqt:function}) for each scaling function, namely $p(k)$, $p(b)$, and $b(k)$, to identify the optimal $p$ value. The insets show the results of plotting Eq.~(\ref{eqt:function}) against various $p$ values. The optimal $p$ value is indicated by a vertical red line in the insets. The shaded region reflects the precision in measuring $p$, based on the fixed $p$ values used in our primary numerical calculations: 0, 0.0001, 0.001, 0.005, 0.01, 0.03, 0.05, 0.15, 0.25, 0.5, 0.75, 0.95, and 1. From Fig.~\ref{fig:real_networks_1} to  Fig.~\ref{fig:real_networks_9}, we present the DC and BC distributions, as well as their correlations, for the remaining 44 real networks. The insets shown in Fig.~\ref{fig:sample_RNs} have been excluded from these figures due to space constraints.\\

Although certain real networks are labeled as directed or weighted based on data collection, we treat them as simple undirected and unweighted networks when identifying their key characteristics by disregarding directionality and considering all non-zero weights to be equal to one. This simplification enables us to uncover universal patterns and principles that may apply across various complex networks, regardless of the specific details of individual networks.\\

Below, we provide a brief overview of the real networks analyzed, focusing on their entities (nodes) and the nature of the connections between them. The numbers in the titles correspond to the indices assigned to these networks, as referenced in the main paper.\\

\textit{1. wiktio. simple — Wiktionary edits (2010)}\\
Three bipartite networks, representing user-page interactions in Wiktionary, were generated for French, German, and English. In these networks, a connection between a user and the page ``simple'' occurs when the user edits that page. The weight of the edges corresponds to the total number of edits made~\cite{wikimedia_dumps}.\\

\textit{2. lkml-thread — Linux kernel mailing list}\\
The network represents user contributions to threads on the Linux kernel mailing list, structured as a bipartite graph. In this network, left nodes correspond to users, and right nodes represent threads. Each edge $(i,j)$ indicates that user $i$ contributed to thread $j$~\cite{kunegis2013konect}.\\

\textit{3. dbp-rcrdlbl — DBpedia artist-label affiliations}\\
Bipartite networks representing the affiliations between artists and the record labels they have performed under, as gathered from Wikipedia data through the DBpedia project. These networks capture the contractual relationships between musicians and their associated record labels~\cite{auer2007dbpedia}.\\

\textit{4. twttr-moscow — Twitter, MoscowAthletics2013}\\
A multiplex network of retweets, mentions, and replies among X (former Twitter) users during the Moscow Athletics event in 2013~\cite{omodei2015characterizing}.\\

\textit{5. pgp-strong — PGP web of trust (2009)}\\
Strongly connected component of the Pretty-Good-Privacy (PGP) web of trust among users, circa November 2009~\cite{richters2011trust}.\\

\textit{6. gowalla — Location-based social network}\\
Location-based online social network, where nodes are accounts and edges are declared friendships~\cite{cho2011friendship}.\\

\textit{7. wiki-talk-ja — Wikipedia talk networks}\\
Interactions among Japanese users of Wikipedia. The nodes represent registered Wikipedia editors, while an edge signifies that user i has posted a message on user j's talk page~\cite{kunegis2013konect}.\\

\textit{8. as-I - CAIDA AS graphs}\\
A network snapshot on 2007-09-17 representing Autonomous System (AS) relationships on the Internet. This was derived using the Serial-1 method, based on RouteViews BGP table snapshots and a set of heuristics.~\cite{caida_as_relationships, leskovec2005graphs}.\\

\textit{9. anybeat - Anybeat social network (2013)}\\
 A snapshot of the Anybeat online social network in 2013, prior to its closure. The nodes symbolize users, and the connections indicate friendships. The direction of the edge $(i, j)$ signifies that user $i$ is following user $j$~\cite{fire2013link}.\\

 \textit{10. wiki-users — Wikipedia user interaction (2011)}\\
 A network derived from interactions between editors of the English language Wikipedia, as derived from the edit histories of 563 wiki pages related to politics~\cite{maniu2011casting}.\\

 \textit{11. wikiquote-tr — Wikiquote edits (2010)}\\
 A bipartite network of users and pages extracted from Wikiquote (in Turkish language), where a user is connected to a page if they have edited that page.~\cite{wikimedia_dumps}.\\

 \textit{12. inploid — Inploid: an online social Q\&A platform}\\
 Inploid is a Turkish social question-and-answer platform where users can follow others and view their questions and answers on the main page. Each user has a reputability score, which is determined by feedback on their questions and answers~\cite{gursoy2018influence}.\\

\textit{13. jdk — Java SE Dev Kit dependencies (1.6.0.7)}\\
A network of class dependencies within the JDK (Java SE Development Kit) version 1.6.0.7. The nodes represent individual classes, and a directed edge signifies that one class depends on another~\cite{kunegis2013konect}.\\

\textit{14. mathoverf. — User interactions on Q\&A websites (2016)}\\
Networks of interactions among users on the Math Overflow online Q\&A site. A directed edge $(i, j)$ signifies that user $i$ commented on an answer provided by user $j$~\cite{paranjape2017motifs}. \\

\textit{15. gentc-mus — Multiplex genetic interactions (2014)}\\
Multiplex networks depicting various types of genetic interactions across different organisms (Mus). The layers correspond to (i) physical, (ii) association, (iii) co-localization, (iv) direct, (v) suppressive, and (vi) additive or synthetic genetic interactions. A directed edge $(i, j)$ indicates that gene $i$ interacts with gene $j$~\cite{de2015muxviz}.\\

\textit{16. dbp-occptin — Person-Occupation Affiliations (DBpedia, 2016)}\\
A bipartite network representing the affiliations between notable individuals and their occupations, sourced from Wikipedia through the DBpedia project~\cite{auer2007dbpedia}. \\

\textit{17. wiki-talk-nds — Wikipedia talk networks}\\
Interactions among Japanese users of Wikipedia. The nodes represent registered Wikipedia editors, while an edge signifies that user i has posted a message on user j's talk page~\cite{kunegis2013konect}.\\

\textit{18. wikinews-tr — Wikipedia news edits (2010)} \\
Two bipartite user-page networks extracted from Wikipedia, about news events (tr). A user connects to a page if that user edited that page~\cite{wikimedia_dumps}. \\

\textit{19. ppi-mouse - MIST protein interaction database (2020)}\\
The Molecular Interaction Search Tool (MIST) is an extensive database of molecular interactions, compiled from multiple primary sources. ``PPI” refers to physical interactions between two or more proteins~\cite{hu2018molecular}.\\

\textit{20. gentc-homo — Multiplex genetic interactions (2014)}\\
Multiplex networks depicting various types of genetic interactions across different organisms (Homo). The layers correspond to (i) physical, (ii) association, (iii) co-localization, (iv) direct, (v) suppressive, and (vi) additive or synthetic genetic interactions. A directed edge (i, j) indicates that gene i interacts with gene j~\cite{de2015muxviz}.\\

\textit{21. topology — Internet AS graph (2004)} \\
A comprehensive view of the Internet's structure at the Autonomous Systems (ASs) level, compiled from various sources such as RouteViews and RIPE BGP trace collectors, route servers, looking glasses, and Internet Routing Registry databases. This snapshot reflects the state of the network around October 2004~\cite{zhang2005collecting}. \\

\textit{22. escorts — Brazilian prostitution network (2010)} \\
A bipartite network of escorts and individuals who purchase sexual services from them in Brazil, derived from a Brazilian online community focused on rating such interactions. The edges indicate instances where sexual services were bought from an escort~\cite{rocha2011simulated}. \\

\textit{23. dbp-country — Person-Country Affiliations (DBpedia, 2016)} \\
A bipartite network representing the affiliations between notable individuals and countries worldwide, derived from Wikipedia through the DBpedia project. The countries include not only current nations but also former countries, empires, kingdoms, and some country-like entities~\cite{auer2007dbpedia}. \\

\textit{24. foursquare - Foursquare global friendships (2013)} \\
A network of user friendships on Foursquare, spanning from April 2012 to September 2013. The `old' and `new' datasets represent two snapshots of the network, taken before and after the check-in data collection period, respectively~\cite{yang2015nationtelescope}. We analysed the `old' one. \\

\textit{25. nmtd-mml — Global nematode–mammal interactions (2018)} \\
A global interaction network of relationships between nematodes and their mammal host species, sourced from the helminthR package and dataset. The nodes are labeled with species-level details for both nematodes and their mammal hosts~\cite{dallas2018gauging}.\\

\textit{26. wordnet — WordNet relationships}\\
A network of English words derived from WordNet, where each node represents a word, and edges indicate various relationships between words, such as synonymy, hypernymy, and meronymy~\cite{fellbaum2010wordnet}. The exact date of extraction from WordNet is unknown. \\

\textit{27. wiktio. or — Wiktionary edits (2010)}\\
Three bipartite networks, representing user-page interactions in Wiktionary, were generated for French, German, and English. In these networks, a connection between a user and the page ``or'' occurs when the user edits that page. The weight of the edges corresponds to the total number of edits made~\cite{wikimedia_dumps}.\\

\textit{28. route-views — Route Views AS graphs} \\
A series of 733 daily network snapshots representing BGP traffic between Autonomous Systems (ASs) on the Internet, sourced from the Oregon Route Views Project. The data spans from November 8, 1997, to January 2, 2000, and was collected by NLANR/MOAT~\cite{leskovec2005graphs} We analysed the  snapshot taken on 2000-01-02. \\

\textit{29. wikibooks-it — Wikipedia book edits (2010)} \\
Two bipartite user-page networks extracted from Wikipedia, about books (the `it' case). A user connects to a page if that user edited that page~\cite{wikimedia_dumps}.\\

\textit{30. mrvl-unvrs — Marvel Universe social network}\\
The Marvel Universe collaboration network, where two Marvel characters are connected if they appear together in the same Marvel comic book~\cite{alberich2002marvel}.\\

\textit{31. academ-edu — Academica.edu (2011)}\\
A snapshot of follower relationships among users of Academia.edu, a platform for academics to share research papers, collected in 2011. In this network, nodes represent users, and a directed edge $(i, j)$ indicates that user $i$ follows user $j$~\cite{fire2014computationally}. \\

\textit{32. email-eu — Email network (EU research inst.)} \\
An anonymized email network from a large European research institution, collected over 18 months from October 2003 to May 2005. Each node represents an email address, and a directed edge between nodes $i$ and $j$ indicates that node $i$ sent at least one message to node $j$~\cite{leskovec2007graph}.\\

\textit{33. chess — Kaggle chess players (2010)}\\
A network of chess players (nodes) representing the outcomes of their matches (edges), detailing game-by-game results among the world’s top chess players. The direction of an edge $(i, j)$ indicates that player $i$ (the white player) played against player $j$ (the black player). Each edge is timestamped (approximately) and is marked with +1 for a win by the white player, 0 for a draw, and -1 for a win by the black player~\cite{chess}.\\

\textit{34. word-adj — Word Adjacency Networks}\\
Directed Network of word adjacency in texts of Spanish language~\cite{milo2004superfamilies}.\\

\textit{35. askubuntu — User interactions on Q\&A websites (2016)} \\
Networks representing interactions among users from Ask Ubuntu online Q\&A platform. A directed edge $(i, j)$ signifies that user $i$ responded to user $j$'s post. For each Q\&A site, four distinct networks are provided based on different edge definitions: (i) user $i$ answered a question from user $j$, (ii) user $i$ commented on a question from user $j$, (iii) user $i$ commented on an answer from user $j$, and (iv) a combined network representing all three types of interactions~\cite{paranjape2017motifs}. We analysed the most general network.\\

\textit{36. enron — Email network (Enron corpus)} \\
The Enron email corpus, consisting of all email communications from the Enron corporation, made public due to legal proceedings. In this network, nodes represent email addresses, and a directed edge from node $i$ to node $j$ indicates that $i$ sent at least one email to $j$. Non-Enron email addresses are also included, but only their interactions with Enron addresses are recorded~\cite{klimt2004enron}.\\

\textit{37. twitter-NY — Twitter, NYClimateMarch2014}\\
A multiplex network of retweets, mentions, and replies among X (former Twitter) users about  the NYClimateMarch2014~\cite{omodei2015characterizing}.\\

\textit{38. lkml-reply — Linux kernel reply network} \\
A network of email replies among addresses from the Linux kernel mailing list. Each node represents an email address, and a directed edge from node $i$ to node $j$ indicates that $i$ replied to an email from $j$~\cite{kunegis2013konect}. The exact date of this snapshot is unknown. \\

\textit{39. proc. Fr — EU national procurement networks (2008-2016)} \\
We analyzed the network ``FR\_2011'' among  234 networks represent the annual national public procurement markets of 26 European countries from 2008-2016, inclusive. Data is sourced from Tenders Electronic Daily (TED), the official procurement portal of the European Union. Nodes with the suffix ``\_$i$" are issuers (sometimes referred to as buyers) of public contracts, for instance public hospitals, ministries, local governments. Nodes with the suffix ``\_$w$" are winners (sometimes called suppliers) of public contracts, generally private-sector firms~\cite{wachs2021corruption}. \\

\textit{40. digg reply — Digg reply network (2008)}\\
Network of replies among users of digg.com. Each node in the network is a digg user, and each directed edge indicates that user $i$ replied to user $j$~\cite{de2009social}.\\

\textit{41. genetic-fly — MIST protein interaction database (2020)
}
The Molecular Interaction Search Tool (MIST) is an extensive database of molecular interactions, compiled from multiple primary sources. Genetic interactions indicate that the effects of mutations in one gene can be modified by mutation of another gene~\cite{hu2018molecular}. \\

\textit{42. as-II - CAIDA AS graphs}\\
The same description as in 8th network. The snapshot has been taken on 2007-11-05~\cite{caida_as_relationships, leskovec2005graphs}.\\

\textit{43. phys.collab. — Multilayer physicist collaborations (2015)} \\
A multiplex network of coauthorships among researchers who have posted preprints on arXiv.org, encompassing all papers up to May 2014. Each layer represents a different category of publication, and the weight of an edge reflects the number of reports coauthored by the researchers. These layers are one-mode projections derived from the underlying author-paper bipartite network~\cite{de2015identifying}.\\

\textit{44. proc. De — EU national procurement networks (2008-2016)} \\
The same description as in 39. We analyzed the network ``DE\_2008''~\cite{wachs2021corruption}. \\

\textit{45. wikibooks-ar — Wikipedia book edits (2010)} \\
Two bipartite user-page networks extracted from Wikipedia, about books (the `ar' case). A user connects to a page if that user edited that page~\cite{wikimedia_dumps}.\\

\textit{46. dbp-location — DBpedia entity-location network}\\
A bipartite network representing the affiliations between named entities from Wikipedia and notable locations, extracted via the DBpedia project. The exact date of this snapshot is unknown~\cite{auer2007dbpedia}.\\

 \textit{47. wikiquote-en — Wikiquote edits (2010)}\\
 A bipartite network of users and pages extracted from Wikiquote (in English language), where a user is connected to a page if they have edited that page.~\cite{wikimedia_dumps}.

\begin{table*}
\caption{Comparison of critical exponents and their corresponding $p$ values across different networks}
\centering
\begin{ruledtabular}
\begin{tabular}{cccccccccc}
 & Network & $\gamma$ & $\delta$ &  $\eta$ & $\frac{2\delta}{\gamma+1}$ & $\frac{\eta(\delta-1)}{\gamma-1}$ & $p_{\gamma}$ & $p_{\delta}$ & $p_{\eta}$ \\ \hline
1 & wiktio. simple~\cite{wikimedia_dumps} &
$1.60\pm0.06$ &
$1.61\pm0.03$ &
$1.09\pm0.06$ &
$1.24\pm0.04$ &
$1.1\pm0.1$ &
$0.03^{+0.02}_{-0.02}$ &
$0.03^{+0.02}_{-0.02}$ &
$0.25^{+0.25}_{-0.10}$ \\
2 & lkml-thread~\cite{kunegis2013konect} &
$2.06\pm0.05$ &
$1.899\pm0.001$ &
$1.1\pm0.1$ &
$1.24\pm0.02$ &
$1.0\pm0.1$ &
$0.50^{+0.25}_{-0.25}$ &
$0.50^{+0.25}_{-0.25}$ &
$0.50^{+0.25}_{-0.25}$ \\
3 & dbp-rcrdlbl~\cite{auer2007dbpedia} &
$2.1\pm0.1$ &
$1.93\pm0.02$ &
$1.17\pm0.05$ &
$1.25\pm0.04$ &
$1.0\pm0.1$ &
$0.50^{+0.25}_{-0.25}$ &
$0.75^{+0.20}_{-0.25}$ &
$0.50^{+0.25}_{-0.25}$ \\
4 & twttr-moscow~\cite{omodei2015characterizing} &
$2.82\pm0.07$ & 
$1.94\pm0.03$ & 
$2\pm1$ & 
$1.02\pm0.02$ & 
$1.0\pm0.6$ & 
$0.95^{+0.05}_{-0.20}$ & 
$0.95^{+0.05}_{-0.20}$ & 
$1.00_{-0.05}$ \\
5 & pgp-strong~\cite{richters2011trust} & $2.92\pm 0.05$ & 
$2.063\pm 0.002$ & 
$1.5\pm0.3$ & 
$1.05\pm0.01$ & 
$0.8\pm0.2$ & 
$1.00_{-0.05}$ & 
$1.00_{-0.05}$ &  
$0.75^{+0.20}_{-0.25}$ \\ 
6 & gowalla~\cite{cho2011friendship} & 
$2.70\pm 0.03$ & 
$2.26\pm0.01$ & 
$1.24\pm0.02$ & 
$1.22\pm0.01$ & 
$0.92\pm 0.03$ & 
$0.95^{+0.05}_{-0.20}$ & 
$1.00_{-0.05}$ &  
$0.50^{+0.25}_{-0.25}$ \\
7 & wiki-talk-ja~\cite{kunegis2013konect} & 
$1.75\pm 0.01$ & 
$1.85\pm0.03$ & 
$1.2\pm0.1$ & 
$1.34\pm0.03$ & 
$1.3\pm 0.1$ & 
$0.15^{+0.10}_{-0.10}$ & 
$0.15^{+0.10}_{-0.10}$ & 
$0.50^{+0.25}_{-0.25}$ \\
8 & as-I~\cite{caida_as_relationships, leskovec2005graphs} &
$1.89\pm0.08$ &
$1.8\pm0.1$ &
$1.28\pm0.04$ &
$1.26\pm0.08$ &
$1.2\pm0.2$ &
$0.25^{+0.25}_{-0.10}$ &
$0.15^{+0.10}_{-0.10}$ &
$0.75^{+0.20}_{-0.25}$ \\
9 & anybeat~\cite{fire2013link} & 
$2.0\pm0.1$ & 
$1.81\pm0.02$ & 
$1.1\pm0.1$ & 
$1.20\pm0.05$ & 
$0.9\pm0.1$ & 
$0.50^{+0.25}_{-0.25}$ & 
$0.15^{+0.10}_{-0.10}$ & 
$0.50^{+0.25}_{-0.25}$ \\ 
10 & wiki-users~\cite{maniu2011casting} &
$2.60\pm0.05$ &
$2.19\pm0.01$ &
$1.4\pm0.2$ &
$1.22\pm0.02$ &
$1.1\pm0.1$ &
$0.95^{+0.05}_{-0.20}$ &
$1.00_{-0.05}$ &
$0.75^{+0.20}_{-0.25}$ \\
11 & wikiquote-tr~\cite{wikimedia_dumps} &
$2.0\pm0.1$ &
$1.87\pm0.03$ &
$1.3\pm0.1$ &
$1.25\pm0.05$ &
$1.1\pm0.1$ &
$0.25^{+0.25}_{-0.10}$ &
$0.25^{+0.25}_{-0.10}$ &
$0.75^{+0.20}_{-0.25}$ \\
12 & inploid~\cite{gursoy2018influence} &
$2.26\pm0.02$ &
$1.85\pm0.04$ &
$1.2\pm0.2$ &
$1.13\pm0.03$ &
$0.8\pm0.2$ &
$0.50^{+0.25}_{-0.25}$ &
$0.15^{+0.10}_{-0.10}$ &
$0.50^{+0.25}_{-0.25}$ \\
13 & jdk~\cite{kunegis2013konect} &
$1.96\pm0.08$ &
$1.7\pm0.1$ &
$1.3\pm0.2$ &
$1.17\pm0.09$ &
$1.0\pm0.3$ &
$0.25^{+0.25}_{-0.10}$ &
$0.05^{+0.10}_{-0.02}$ &
$0.75^{+0.20}_{-0.25}$ \\
14 & mathoverf.~\cite{paranjape2017motifs} &
$1.78\pm0.04$ &
$1.82\pm0.02$ &
$1.4\pm0.1$ &
$1.31\pm0.02$ &
$1.5\pm0.1$ &
$0.25^{+0.25}_{-0.10}$ &
$0.15^{+0.10}_{-0.10}$ &
$0.75^{+0.20}_{-0.25}$ \\
15 & gentc-mus~\cite{de2015muxviz} &
$2.1\pm0.1$ &
$1.97\pm0.05$ &
$1.2\pm0.4$ &
$1.26\pm0.05$ &
$1.1\pm0.4$ &
$0.50^{+0.25}_{-0.25}$ &
$1.00_{-0.05}$ &
$0.50^{+0.25}_{-0.25}$ \\
16 & dbp-occptin~\cite{auer2007dbpedia} &
$1.73\pm0.03$ &
$1.85\pm0.03$ &
$1.01\pm0.07$ &
$1.36\pm0.03$ &
$1.2\pm0.1$ &
$0.15^{+0.10}_{-0.10}$ &
$0.15^{+0.10}_{-0.10}$ &
$0.05^{+0.10}_{-0.02}$ \\
17 & wiki-talk-nds~\cite{kunegis2013konect} &
$1.64\pm0.06$ &
$1.5\pm0.1$ &
$1.2\pm0.3$ &
$1.1\pm0.1$ &
$1.0\pm0.4$ &
$0.05^{+0.10}_{-0.02}$ &
$0.01^{+0.02}_{-0.005}$ &
$0.50^{+0.25}_{-0.25}$ \\
18 & wikinews-tr~\cite{wikimedia_dumps} &
$1.7\pm0.1$ &
$1.74\pm0.03$ &
$1.17\pm0.08$ &
$1.27\pm0.06$ &
$1.2\pm0.2$ &
$0.15^{+0.10}_{-0.10}$ &
$0.05^{+0.10}_{-0.02}$ &
$0.50^{+0.25}_{-0.25}$ \\
19 & ppi-mouse~\cite{hu2018molecular} &
$2.33\pm0.02$ &
$1.942\pm0.006$ &
$1.40\pm0.09$ &
$1.166\pm0.007$ &
$0.99\pm0.06$ &
$0.75^{+0.20}_{-0.25}$ &
$0.95^{+0.05}_{-0.20}$ & 
$0.75^{+0.20}_{-0.25}$ \\
20 & gentc-homo~\cite{de2015muxviz} &
$2.12\pm0.02$ &
$1.90\pm0.02$ &
$1.1\pm0.2$ &
$1.21\pm0.02$ &
$0.9\pm0.2$ &
$0.50^{+0.25}_{-0.25}$ &
$0.25^{+0.25}_{-0.10}$ &
$0.50^{+0.25}_{-0.25}$ \\
21 & topology~\cite{zhang2005collecting} &
$1.83\pm0.02$ &
$1.789\pm0.009$ &
$1.1\pm0.2$ &
$1.27\pm0.01$ &
$1.1\pm0.2$ &
$0.25^{+0.25}_{-0.10}$ &
$0.15^{+0.10}_{-0.10}$ &
$0.50^{+0.25}_{-0.25}$ \\
22 & escorts~\cite{rocha2011simulated} &
$2.5\pm0.1$ &
$2.26\pm0.01$ &
$1.17\pm0.09$ &
$1.28\pm0.04$ &
$1.0\pm0.1$ &
$0.75^{+0.20}_{-0.25}$ &
$1.00_{-0.05}$ &
$0.50^{+0.25}_{-0.25}$ \\
23 & dbp-country~\cite{auer2007dbpedia} &
$1.49\pm0.03$ &
$1.85\pm0.02$ &
$1.05\pm0.04$ &
$1.48\pm0.02$ &
$1.8\pm0.1$ &
$0.0001^{+0.0009}_{-0.0001}$ &
$0.15^{+0.10}_{-0.10}$ &
$0.15^{+0.10}_{-0.10}$ \\
24 & foursquare~\cite{yang2015nationtelescope} &
$2.35\pm0.09$ &
$2.111\pm0.004$ &
$1.4\pm0.3$ &
$1.26\pm0.04$ &
$1.1\pm0.2$ &
$0.75^{+0.20}_{-0.25}$ &
$1.00_{-0.05}$ &
$0.75^{+0.20}_{-0.25}$ \\
25 & nmtd-mml~\cite{dallas2018gauging} &
$2.28\pm0.09$ &
$2.12\pm0.02$ &
$1.2\pm0.2$ &
$1.29\pm0.04$ &
$1.1\pm0.2$ &
$0.50^{+0.25}_{-0.25}$ &
$1.00_{-0.05}$ &
$0.75^{+0.20}_{-0.25}$ \\
26 & wordnet~\cite{fellbaum2010wordnet} & $2.58\pm0.03$ & 
$2.13\pm0.05$ & 
$1.7\pm0.2$ & 
$1.19\pm0.03$ & 
$1.2\pm0.2$ & 
$0.95^{+0.05}_{-0.20}$ & 
$1.00_{-0.05}$ & 
$0.95^{+0.05}_{-0.20}$ \\
27 & wiktio. or~\cite{wikimedia_dumps} &
$1.5\pm0.1$ &
$1.53\pm0.04$ &
$1.06\pm0.02$ &
$1.25\pm0.06$ &
$1.2\pm0.3$ &
$0.0001^{+0.0009}_{-0.0001}$ &
$0.01^{+0.02}_{-0.005}$ &
$0.15^{+0.10}_{-0.10}$ \\
28 & route-views~\cite{leskovec2005graphs} &
$2.06\pm0.07$ &
$1.85\pm0.03$ &
$1.30\pm0.05$ &
$1.21\pm0.03$ &
$1.04\pm0.09$ &
$0.50^{+0.25}_{-0.25}$ &
$0.15^{+0.10}_{-0.10}$ &
$0.75^{+0.20}_{-0.25}$ \\
29 & wikibooks-it~\cite{wikimedia_dumps} &
$2.162\pm0.009$ &
$1.85\pm0.01$ &
$1.3\pm0.1$ &
$1.170\pm0.007$ &
$0.9\pm0.1$ &
$0.50^{+0.25}_{-0.25}$ &
$0.15^{+0.10}_{-0.10}$ &
$0.75^{+0.20}_{-0.25}$ \\
30 & mrvl-unvrs~\cite{alberich2002marvel} &
$2.26\pm0.06$ &
$1.97\pm0.03$ &
$1.4\pm0.1$ &
$1.21\pm0.03$ &
$1.1\pm0.1$ &
$0.50^{+0.25}_{-0.25}$ &
$1.00_{-0.05}$ &
$0.75^{+0.20}_{-0.25}$ \\
31 & academ-edu~\cite{fire2014computationally} &
$2.61\pm0.04$ &
$2.13\pm0.01$ &
$1.3\pm0.1$ &
$1.18\pm0.02$ &
$0.93\pm0.08$ &
$0.95^{+0.05}_{-0.20}$ &
$1.00_{-0.05}$ &
$0.75^{+0.20}_{-0.25}$ \\
32 & email-eu~\cite{leskovec2007graph} &
$1.89\pm0.05$ &
$1.8\pm0.2$ &
$1.4\pm0.2$ &
$1.3\pm0.1$ &
$1.3\pm0.4$ &
$0.25^{+0.25}_{-0.10}$ &
$0.15^{+0.10}_{-0.10}$ &
$0.75^{+0.20}_{-0.25}$ \\
33 & chess~\cite{chess} &
$1.7\pm0.2$ &
$1.78\pm0.03$ &
$1.1\pm0.1$ &
$1.32\pm0.08$ &
$1.3\pm0.3$ &
$0.15^{+0.10}_{-0.10}$ &
$0.15^{+0.10}_{-0.10}$ &
$0.50^{+0.25}_{-0.25}$ \\
34 & word-adj~\cite{milo2004superfamilies} & 
$2.06\pm0.07$ & 
$1.876\pm0.006$ & $1.31\pm0.09$ & $1.23\pm0.03$ & $1.1\pm0.1$ & $0.50^{+0.25}_{-0.25}$ & $0.25^{+0.25}_{-0.10}$ & $0.75^{+0.20}_{-0.25}$ \\
35 & askubuntu~\cite{paranjape2017motifs} & 
$1.97\pm0.09$ &
$1.907\pm0.002$ &
$1.19\pm0.06$ &
$1.28\pm0.04$ &
$1.1\pm0.1$ &
$0.25^{+0.25}_{-0.10}$ & 
$0.50^{+0.25}_{-0.25}$ & 
$0.50^{+0.25}_{-0.25}$ \\
36 & enron~\cite{klimt2004enron} & $2.22\pm 0.02$ & 
$1.95\pm 0.02$ & 
$1.5\pm 0.2$ & 
$1.21\pm0.01$ & 
$1.2\pm0.1$ & 
$0.50^{+0.25}_{-0.25}$ & 
$1.00_{-0.05}$ & 
$0.95^{+0.05}_{-0.20}$  \\ 
37 & twitter-NY~\cite{omodei2015characterizing} &
$2.4\pm0.1$ &
$1.99\pm0.04$ &
$1.84\pm0.03$ &
$1.17\pm0.05$ &
$1.3\pm0.1$ &
$0.75^{+0.20}_{-0.25}$ &
$1.00_{-0.05}$ & 
$1.00_{-0.05}$ \\
38 & lkml-reply~\cite{kunegis2013konect} &
$1.88\pm0.06$ &
$2.05\pm0.05$ &
$1.26\pm0.04$ &
$1.42\pm0.04$ &
$1.5\pm0.1$ &
$0.25^{+0.25}_{-0.10}$ &
$1.00_{-0.05}$ &
$0.75^{+0.20}_{-0.25}$ \\
39 & proc. Fr~\cite{wachs2021corruption} &
$2.05\pm0.09$ &
$2.01\pm0.03$ &
$1.2\pm0.1$ &
$1.31\pm0.04$ &
$1.1\pm0.1$ &
$0.50^{+0.25}_{-0.25}$ &
$1.00_{-0.05}$ &
$0.50^{+0.25}_{-0.25}$ \\
40 & digg reply~\cite{de2009social} &
$2.8\pm0.2$ &
$2.12\pm0.02$ &
$1.4\pm0.2$ &
$1.11\pm0.05$ &
$0.9\pm0.2$ &
$0.95^{+0.05}_{-0.20}$ &
$1.00_{-0.05}$ &
$0.75^{+0.20}_{-0.25}$ \\
41 & genetic-fly~\cite{hu2018molecular} &
$1.91\pm0.06$ &
$2.10\pm0.03$ &
$1.47\pm0.07$ &
$1.44\pm0.04$ &
$1.8\pm0.2$ &
$0.25^{+0.25}_{-0.10}$ &
$1.00_{-0.05}$ &
$0.75^{+0.20}_{-0.25}$ \\
42 & as-II~\cite{caida_as_relationships, leskovec2005graphs} &
$2.14\pm0.01$ &
$1.82\pm0.02$ &
$1.23\pm0.08$ &
$1.16\pm0.01$ &
$0.89\pm0.06$ &
$0.50^{+0.25}_{-0.25}$ &
$0.15^{+0.10}_{-0.10}$ &
$0.50^{+0.25}_{-0.25}$ \\
43 & phys.collab.~\cite{de2015identifying} &
$2.4\pm0.5$ &
$2.10\pm0.02$ &
$1.9\pm0.1$ &
$1.3\pm0.2$ &
$1.6\pm0.6$ &
$0.75^{+0.20}_{-0.25}$ &
$1.00_{-0.05}$ &
$1.00_{-0.05}$ \\
44 & proc. De~\cite{wachs2021corruption} &
$2.4\pm0.2$ &
$2.01\pm0.06$ &
$1.2\pm0.2$ &
$1.18\pm0.08$ &
$0.9\pm0.2$ &
$0.75^{+0.20}_{-0.25}$ &
$1.00_{-0.05}$ &
$0.50^{+0.25}_{-0.25}$ \\
45 & wikibooks-ar~\cite{wikimedia_dumps} &
$1.6\pm0.2$ &
$1.87\pm0.05$ &
$1.1\pm0.1$ &
$1.4\pm0.1$ &
$1.5\pm0.5$ &
$0.05^{+0.10}_{-0.02}$ &
$0.25^{+0.25}_{-0.10}$ &
$0.25^{+0.25}_{-0.10}$ \\
46 & dbp-location~\cite{auer2007dbpedia} &
$2.08\pm0.08$ &
$2.11\pm0.03$ &
$1.2\pm0.1$ &
$1.37\pm0.04$ &
$1.3\pm0.2$ &
$0.50^{+0.25}_{-0.25}$ &
$1.00_{-0.05}$ &
$0.50^{+0.25}_{-0.25}$ \\
47 & wikiquote-en~\cite{wikimedia_dumps} &
$2.16\pm0.01$ &
$2.045\pm0.003$ &
$1.3\pm0.1$ &
$1.293\pm0.005$ &
$1.1\pm0.1$ &
$0.50^{+0.25}_{-0.25}$ &
$1.00_{-0.05}$ &
$0.75^{+0.20}_{-0.25}$ \\
\end{tabular}
\end{ruledtabular}
\label{tab:network_comparison}
\end{table*}

\end{document}